\documentclass[preprint,12pt]{elsarticle}

\usepackage{amssymb}
\usepackage{rotating,booktabs}
\usepackage{color}

\journal{Materials Chemistry and Physics}

\begin{document}

\begin{frontmatter}



\title{\emph{Ab-initio} study of structural, vibrational and optical properties of solid oxidizers}


\author{N. Yedukondalu and G. Vaitheeswaran}
\address{Advanced Centre of Research in High Energy Materials (ACRHEM), University of Hyderabad, Prof. C. R. Rao Road, Gachibowli, Hyderabad-500046, Telangana, India. E-mail: gvsp@uohyd.ernet.in}

\begin{abstract}
We report the structural, elastic and vibrational properties of five ionic-molecular solid oxidizers MNO$_3$ (M = Li, Na, K) and MClO$_3$ (M = Na, K). By treating long range electron-correlation effects, dispersion corrected method leads to more accurate predictions of structural properties and phase stability of KNO$_3$ polymorphs. The obtained elastic moduli show soft nature of these materials and are consistent with Ultrasonic Pulse Echo measurements. We made a complete assignment of vibrational modes which are in good accord with available experimental results. From calculated IR and Raman spectra, it is found that the vibrational frequencies show a red-shift from Li $\rightarrow$ Na $\rightarrow$ K (Na $\rightarrow$ K) and N $\rightarrow$ Cl for nitrates (chlorates) due to increase in mass of metal and non-metal atoms, respectively. The calculated electronic structure using recently developed Tran-Blaha modified Becke-Johnson potential show that the materials are wide band gap insulators with predominant ionic bonding between M$^+$ (metal) and NO$_3^-$/ClO$_3^-$ ions and covalent bonding (N-O and Cl-O) within nitrate and chlorate anionic group. From the calculated optical spectra, we observe that electric-dipole transitions are due to nitrate/chlorate group below 20 eV and cationic transitions occur above 20 eV.  The calculated reflectivity spectra are consistent with the available experimental spectra.

\end{abstract}

\begin{keyword}
\emph{ab-initio} calculations, elastic properties, insulators, optical properties



\end{keyword}

\end{frontmatter}


\section{INTRODUCTION}
Pyrotechnic formulations consist of physically intimate mixtures of various combinations of fuels and oxidizers. Fuel is the main burning ingredient in pyrotechnic formulations. The choice of fuels is very wide, ranging from metallic to non-metallic elements and binary compounds to various types of carbonaceous materials, while substances containing oxygen as well as fluorine or chlorine are used as oxidizing agents.\cite{Hofmann} Oxidizers are oxygen-rich ionic solids that decompose at moderate-to-high temperatures by liberating oxygen and composed of nitrates, chlorates, perchlorates, peroxides, chromates, and dinitramides.\cite{Koch, Venkatachalam,Meyer,Vandel,Steinhauser} The oxidizer is a major ingredient of composite propellants and constitutes more than 70$\%$ (by weight) of the propellant. However, the concentration of oxygen within an explosive or oxidizer is represented by a term known as oxygen balance (OB), which is an important parameter for optimizing its application as an explosive or oxidizer. Since most of the secondary explosives including TNT (-74$\%$), HNS (-67.6$\%$), HMX (-21.62$\%$), and RDX (-21.6$\%$) etc., have oxygen deficiency (or negative OB)\cite{Meyer}, it is necessary to bind an oxidizer with secondary explosive to complete the combustion process to form its decomposition products by releasing huge (maximum) amount of energy.
\par Alkali metal based oxidizers are highlighted for non-toxicity (no heavy metal pollution) which leads to eco-friendly decomposition products and also find applications in composite propellants\cite{Meyer,Agrawal,Sutton,Kubota} due to their high positive OB. The alkali metal nitrates and chlorates are efficient oxidizers with positive OB values: LiNO$_3$ (58.1$\%$), NaNO$_3$ (47.1$\%$), KNO$_3$ (39.6$\%$), NaClO$_3$ (45.1$\%$) and KClO$_3$ (39.2$\%$)\cite{Meyer}. LiNO$_3$ is a colorless salt which finds applications in the production of red-colored fire works, flares, submarines and space craft for absorbing excess of CO$_2$ in air. NaNO$_3$ and KNO$_3$ are white crystalline solids and are extensively used in fertilizers, smoke bombs, solid rocket propellants.\cite{Meyer} NaNO$_3$ is used as a raw material whereas KNO$_3$ is a major ingredient (75$\%$) in the production of gunpowder.\cite{Steinhauser} In addition, nitrates also serve as primers which are used to initiate other explosives or blasting agents. Besides their application as oxidizers and primers, NaNO$_3$ and KNO$_3$ have recently attracted much interest towards detection and identification of explosive materials by using Tera-Hertz spectroscopy.\cite{Witko,Fu} The chlorate based propellants are more efficient than traditional black powder propellants. The presence of an oxidizer in an explosive increases the gaseous products in the decomposition mechanism. The decomposition process of the investigated oxidizers is as follows:\cite{Johnson,Babar}
\begin{center}
4LiNO$_3$ $\rightarrow$ 2Li$_2$O + 4NO$_2$ + O$_2$ \\
2NaNO$_3$ $\rightarrow$ 2NaNO$_2$ + O$_2$ \\
2KNO$_3$ $\rightarrow$ 2KNO$_2$ + O$_2$ \\
2NaClO$_3$ $\rightarrow$ 2NaCl + 3O$_2$ \\
2KClO$_3$ $\rightarrow$ 2KCl + 3O$_2$ \\
\end{center}
Thermal and kinetic decomposition studies on ammonium and potassium based oxidizers explore thermal stability, kinetics, sensitivity and the amount of oxygen provided by the oxidizers.\cite{Babar} Thermal decomposition studies on composite materials KClO$_3$-HMX(RDX) have proved that energy release of mixtures exceeds 40$\%$ (10$\%$) to that of pure HMX (RDX).\cite{Liao,Dong}
 Decomposition reactions of (Na/K)ClO$_3$ were investigated by X-ray photo electron spectroscopy.\cite{Park} Several experimental studies have been reported on structural\cite{Wu,Gonschorek,Nimmo,Adiwidjaja,Abrahams,Danielsen} mechanical\cite{Haussuhl,Michard,Viswanathan}, IR and Raman spectroscopy\cite{RamaRao,Ramdas1,Khanna,Nedungadi,Ismail,French,Ferraro,Brooker,kumari,Ramdas2,Nakagawa,James,Rousseau,Liu,Brooker1,Hartwig,Bates,Brooker2,Heyns} and optical\cite{Neufeld,Vorobeva,Zhuravlvev,Burkov} properties of the bulk crystals. A considerable amount of theoretical work has been devoted to understand the phase stability\cite{Aydinol} mechanical\cite{Belomestnykh} electronic structure of bulk and surfaces\cite{Aydinol,Zhuravlev1,Zhuravlev2,Zhuravlev3,Zhuravlev4,Zhuravlev5,Lovvik,Pradeep} and optical\cite{Zhuravlev4} properties of these materials using Hatree-Fock (HF) and standard Density Functional Theory (DFT) methods. Despite of several reports on these materials in the literature, they have apparently not succeeded in predicting the structural, phase stability and electronic properties due to a poor description of the weak dispersive interactions and under-/over-estimation of band gap using standard DFT/HF methods. To address these issues, in the present study, we investigate the structural, elastic and vibrational properties using dispersion corrected DFT-D2 method and recently developed semi-local potential of Tran-Blaha modified Becke Johnson (TB-mBJ) approach to calculate the electronic structure and optical properties of the solid oxidizing materials. The rest of the article is organized as follows: in the next section, we briefly describe the methodology and details of our calculations. Results and discussion concerning the structural, elastic, vibrational, electronic and optical properties of the five investigated compounds. Finally, in the last section, we summarize the conclusions of the present study.

\section{METHODOLOGY}
Plane-wave pseudopotential (PW-PP) approach has been used to perform the first-principles calculations of the five inorganic solid oxidizers using CAmbridge Series of Total Energy Package (CASTEP)\cite{Payne} based on DFT. Electron-ion interactions are treated with ultra soft pseudo potentials (USPP) \cite{Vanderbilt} to calculate the structural and elastic properties, whereas norm conserving\cite {Troullier} pseudo potentials (NCPP) to calculate the structural and vibrational properties of the investigated compounds. The local density approximation (LDA)\cite{Ceperley,Perdew} and generalized gradient approximation (GGA)\cite{Burke} were used to treat electron-electron interactions. The Broyden-Fletcher-Goldfarb-Shanno (BFGS) minimization scheme\cite{Almlof} has been used for structural relaxation. The convergence criteria for structural optimization was set to ultra-fine quality with a kinetic energy cut-off of 750 eV and k-mesh of 6$\times$6$\times$6 for LiNO$_3$ and NaNO$_3$, 5$\times$3$\times$4 for KNO$_3$, 4$\times$4$\times$4 for NaClO$_3$, 6$\times$5$\times$4 for KClO$_3$ according to the Monkhorst-Pack grid scheme.\cite{Monkhorst} The self-consistent energy convergence was set to 5.0$\times$10$^{-6}$ eV/atom. The convergence criterion for the maximum force between atoms was 0.01 eV/$\AA$. The maximum displacement and stress were set to be 5.0$\times$10$^{-4}\AA$ and 0.02 GPa, respectively.

\par Semi-empirical dispersion correction method (DFT-D2) has been used to treat dispersive interactions which was developed by Grimme\cite{Grimme} and implemented through PBE-GGA functional. Here is the total energy after inclusion of dispersion correction is given by
\begin{equation}
E_{DFT+D} = E_{DFT} + E_{disp}
\end{equation}
where E$_{DFT}$ is the self-consistent Kohn-Sham energy, E$_{disp}$ is empirical dispersion correction energy given by
\begin{equation}
E_{disp} = -S_6\sum\limits_{i \textless j}\frac{C_{ij}}{R_{ij}^6}f_{damp}(R_{ij})
\end{equation}
where S$_6$ is global scaling factor that only depends on the density functional used. C$_{ij}$ denotes the dispersion coefficient for the pair of i$^{th}$ and j$^{th}$ atoms that depends on the chemical species, and R$_{ij}$ is an inter-atomic distance. f$_{damp}$ = $\frac{1}{1+e^{-d(R_{ij}/R_0-1)}}$ is a damping function which is necessary to avoid divergence for small values of R$_{ij}$ and R$_0$ is the sum of atomic van der Waals (vdW) radii.

The standard DFT functionals LDA \cite{Ceperley} and GGA \cite{Burke} are capable of determining the ground state properties of solids, which are in reasonable agreement with the experiments. However, the excited state properties like band gap are severely underestimated by 30-40\% when compared to experiments for semiconductors and insulators or even absent for semiconductors \cite{heyd} due to derivative discontinuity of the exchange correlation functional \cite{perdew2,sham}. The Kohn-Sham (KS) band gap is defined as the difference between valence band maximum (VBM) and conduction band minimum (CBM) of the energy eigen value spectrum. While the experimental band gap is measured from the difference between ionization potential (I) and electron affinity (A) \emph{i.e} I-A. Several methods such as LDA+U, LDA+DMFT, hybrid fuctionals and GW approximation have been proposed to get reliable energy band gaps compared to experiments. Unfortunately these methods are computationally very expensive. Therefore, in the present work, Engel Vosko (EV)-GGA \cite{engel} and Tran Blaha modified Becke Johnson (TB-mBJ) potential \cite{peter2} have been used to calculate the band gap of the investigated materials. However, the later one is computationally inexpensive and it provides relaiable energy band gaps as comparable with more sophisticated methods within the KS frame work. This semi local TB-mBJ potential is obtained from simple modification of Becke-Johnson (BJ) potential \cite{becke} by the introduction of local parameter {\bf c} into BJ potential leading to TB-mBJ potential as

\begin{equation}
\nu_{\emph{x}, \sigma}^{mBJ}(r) = {\bf c}\nu_{\emph{x}, \sigma}^{BR}(r) + (3{\bf c}-2)\frac{1}{\pi}\sqrt{\frac{5}{12}}\sqrt{\frac{2t_\sigma(r)}{\rho_\sigma(r)}} \\
\end{equation}

where $\rho_{\sigma}$ = $\Sigma_{i=1}^{N_{\sigma}} |\psi_{i,\sigma}|^2$ is the electron density, $t_{\sigma}$ = $\frac{1}{2}\Sigma_{i=1}^{N_{\sigma}}\bigtriangledown \psi^*_{i,\sigma}\textbf{.}\bigtriangledown \psi_{i,\sigma}$ is the kinetic energy density, $\nu_{\emph{x}, \sigma}^{BR}(r)$ is the Becke-Roussel potential, and {\bf c}  = $\alpha$ + $\beta\large{(}\frac{1}{V_{cell}}\large{\int_{cell}}\frac{|\bigtriangledown\rho(r^{\prime})|}{\rho(r^{\prime})}d^3r^{\prime}\large{)}$, where $\alpha$, $\beta$ are two free parameters and $V_{cell}$ is the unit cell volume.  Recently, Koller \emph{et al}\cite{koller1,koller2} reported the merits and limitations of the TB-mBJ and also they made an attempt to improve the performance of the original TB-mBJ potential by optimizing the local parameter {\bf c} in above equation 3. So far, several groups have been using the TB-mBJ potential for the calculation of electronic structure and optical properties of diverse materials \cite{singh2,camargo,jiang} and conformed that TB-mBJ band gaps are improved for a wide range of materials (see Fig. 1 from Ref. \cite{dixit}) over the standard DFT functionals. The accurate prediction of band gaps using TB-mBJ potential for diverse materials motivated us to use this potential for the oxidizing materials. This semi local potential is implemented through WIEN2k package \cite{blaha}. In order to achieve energy eigenvalue convergence, the wave functions in the interstitial region were expanded in plane waves with a cut-off K$_{max}$ = 8/R$_{MT}$, where R$_{MT}$ is the smallest muffin-tin sphere while the charge density was Fourier expanded up to G$_{max}$ = 14. To calculate the optical spectra one has to use a denser K-mesh in the Brillouin zone because the matrix element changes more rapidly within the Brillouin zone than the electronic energies themselves. Therefore, it requires a denser K-mesh to integrate this property accurately than for a normal self consistent calculation and hence we have used 14x14x14 for (Li/Na)NO$_3$ and NaClO$_3$, 15x18x10 for KNO$_3$ and 17x14x11 for KClO$_3$ for optical properties calculation.

\section{RESULTS AND DISCUSSION}
\subsection{Crystal structure and ground state properties}
\begin{figure}
\centering
\includegraphics[height = 5.5in, width=6.5in]{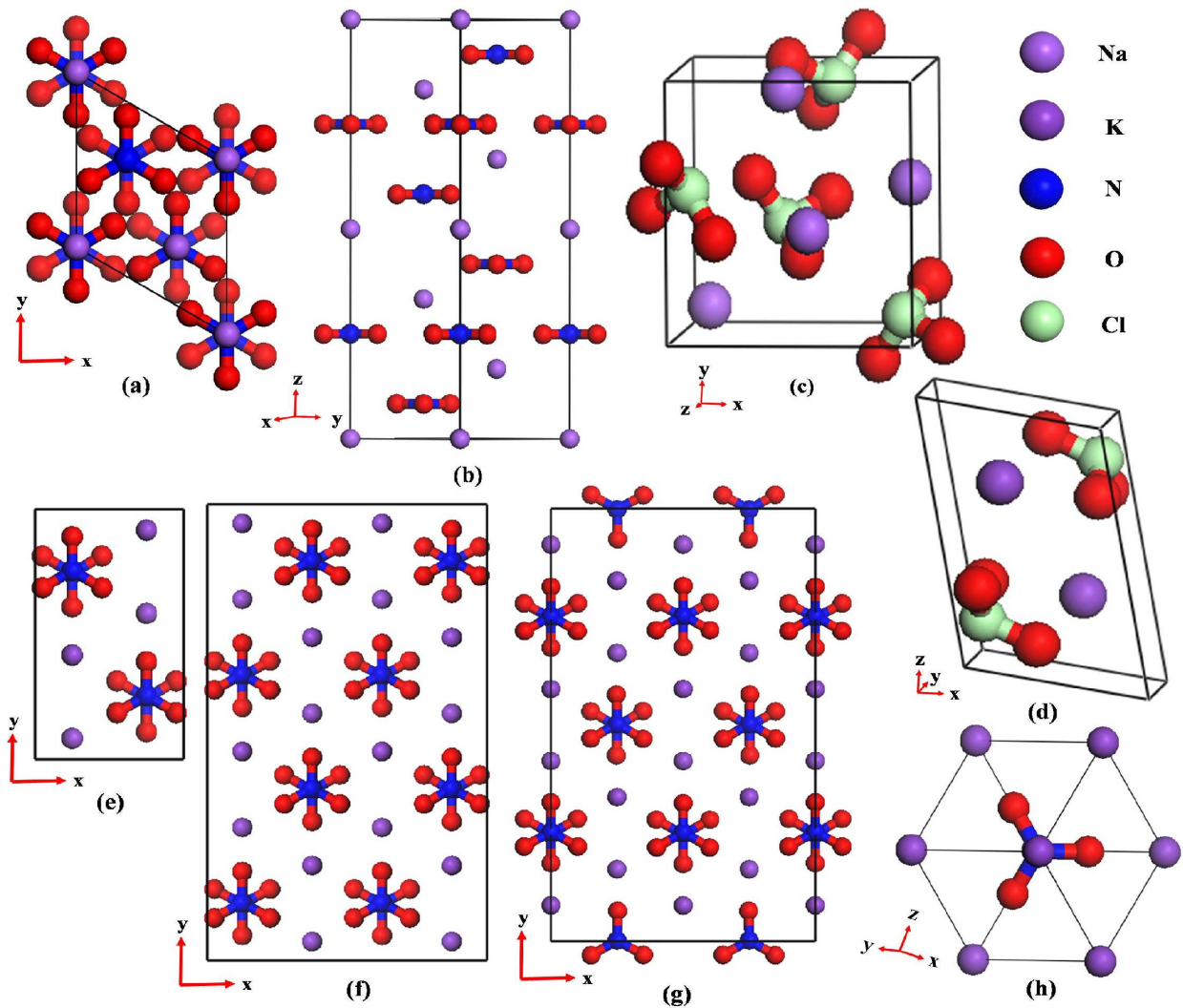}
\caption{\label{crystal} Crystal structures of (a, b) NaNO$_3$, (c) NaClO$_3$, (d) KClO$_3$, and (e, f) unit cell and 2 $\times$ 2 $\times$ 1 super cell of $\alpha$-phase, (g) unit cell of super structure (Cmc2$_1$), (h) $\gamma$-phase of KNO$_3$ polymorphs.}
\label{crystal}
\end{figure}
At ambient conditions, alkali metal nitrates such as LiNO$_3$\cite{Wu} and NaNO$_3$\cite{Gonschorek} crystallize in the trigonal structure having space group $R\bar{3}$c with Z=6 and isomorphic with calcite structure. The alkali metal chlorates namely NaClO$_3$\cite{Abrahams} and KClO$_3$\cite{Danielsen} crystallize in the primitive cubic ($P2_13$, Z=4) and monoclinic ($P2_1/m$, Z=2) structures, respectively as shown in figure 1\ref{crystal}. In order to predict elastic and vibrational properties, we first performed full structural optimization to obtain equilibrium crystal structures of the above mentioned compounds by taking experimental structures as an input. The obtained volumes are underestimated by $\sim$ 9-15$\%$ within LDA, overestimated $\sim$ 6-11.5$\%$ within PBE-GGA when compared to the experimental data.\cite{Wu,Gonschorek,Abrahams,Danielsen} This clearly shows that the obtained unit cell parameters using standard LDA/GGA functionals show a large deviation from the experimental data due to weak dispersive interactions present in these materials. The alkali metal nitrates and chlorates are analogous to simple alkali halides but the presence of planar nitrate and trigonal chlorate anions make the crystal structure more complex. Also these materials are similar to alkali metal azides with a difference in the arrangement of atoms in an anionic group such that nitrates, chlorates and azides possesses planar, trigonal and linear anions, respectively. From our earlier work, we found that the dispersive interactions play a major role in predicting the structural and vibrational properties of ionic alkali metal nitrates and azides at ambient as well as at high pressure.\cite{konda,ramesh1,ramesh2,ramesh3} Apart from this, DFT-D2 method is successful in explaining the phase stability by accounting for the missing dispersive interactions in ionic cesium halides.\cite{Zhang} Hence, in the present work, we have also used the DFT-D2 method to calculate the structural properties of the investigated compounds and the obtained equilibrium volumes differ by $\sim$ 0.56-2.9$\%$ when compared to experimental data.\cite{Wu,Gonschorek,Abrahams,Danielsen} As seen in Table \ref{str1}, the relative errors with DFT-D2 method are small, which clearly show the success of the method in reproducing the structural properties of these oxidizing materials over standard DFT functionals.

\begin{table}
\centering
\caption{Calculated lattice constants (a, b, c, in $\AA$), lattice angle ($\beta$, in $^o$) and volume (V, in $\AA^3$) of LiNO$_3$, NaNO$_3$, NaClO$_3$, and KClO$_3$ using LDA, GGA and dispersion corrected (DFT-D2) methods within the NCPP approach along with experimental data. The relative errors (in $\%$) with respect to (w. r .t) experiments are given in parenthesis, here '-' and '+' signs indicate
underestimation and overestimation of the calculated values w. r .t experiments.}
{\resizebox{0.98\textwidth}{!}{
\begin{tabular}{cccccc}
\hline Compound        & Parameter &      CA-PZ         &      PBE         &     DFT-D2        &  Expt.         \\ \hline
LiNO$_3$               &   a       &   4.655 (-0.79)    &  4.771 (+1.68)   &  4.731 (+0.83)    &  4.692$^a$     \\
                       &   c       &   14.500 (-4.70)   &  15.479 (+1.74)  &  15.173 (-0.28)   &  15.215$^a$    \\
                       &   V       &   272.17 (-6.17)   &  305.23 (+5.22)  &  294.12 (+1.39)   &  290.08$^a$    \\
NaNO$_3$               &   a       &   5.041 (-0.57)    &  5.183 (+2.23)   &  5.167 (+1.91)    &  5.070$^b$     \\
                       &   c       &   15.609 (-7.21)   &  17.206 (+2.28)  &  16.665 (-0.93)   &  16.822$^b$    \\
                       &   V       &   343.51 (-8.21)   &  400.26 (+6.88)  &  385.34 (+2.90)   &  374.49$^b$    \\
NaClO$_3$              &   a       &   6.389 (-2.84)    &  6.721 (+2.20)   &  6.623 (+0.72)    &  6.576$^c$     \\
                       &   V       &   260.79 (-8.29)   &  303.56 (+6.76)  &  290.45 (+2.14)   &  284.35$^c$    \\
KClO$_3$               &   a       &   4.461 (-4.21)    &  4.733 (+1.63)   &  4.734 (+1.66)    &  4.6569$^d$    \\
                       &   b       &   5.418 (-3.09)    &  5.722 (+2.34)   &  5.555 (-0.64)    &  5.5909$^d$    \\
                       &   c       &   6.786 (-4.41)    &  7.249 (+2.11)   &  7.045 (-0.76)    &  7.0991$^d$    \\
                       & $\beta$   &   111.66 (+1.83)   &  107.11 (-2.32)  &  110.87 (+1.11)   &  109.65$^d$    \\
                       &   V       &   154.41 (-11.3)   &  187.63 (+7.79)  &  173.10 (-0.56)   &  174.07$^d$    \\ \hline
\end{tabular}}}
\\ $^a$Ref.\cite{Wu},
$^b$Ref.\cite{Gonschorek},
$^c$Ref.\cite{Abrahams},
$^d$Ref.\cite{Danielsen},
\label{str1}
\end{table}

\par In addition, we also attempted to study the phase stability of KNO$_3$ polymorphs at ambient pressure. It is well known from the literature, KNO$_3$ crystallizes in the primitive orthorhombic structure which belongs to $Pmcn$ space group with Z=4 (also called $\alpha$-phase or phase II) at room temperature.\cite{Nimmo} A super-structure of $\alpha$-KNO$_3$ has been reported by Adiwidjaja et al\cite{Adiwidjaja} with crystal symmetry Cmc2$_1$ which consists of a 2 $\times$ 2 $\times$ 1 super cell of the $Pmcn$ unit cell with Z = 16. When heated the $\alpha$-phase transforms to $\beta$ phase (I) with disordered nitrate anions at around 128 $^o$C,\cite{Nimmo2} and it passes through the ferroelectric $\gamma$-phase (III) between 124 and 110 $^o$C upon cooling.\cite{Nimmo3} The crystal structures of KNO$_3$ polymorphs are shown in figure \ref{crystal}. The previous calculations show that the volume of $\alpha$-phase is severely underestimated around 34.7 $\%$\cite{Aydinol} and 15.3 $\%$\cite{Lu} within LDA and the volume of Cmc2$_1$ super structure is overestimated by 4.5 $\%$ within PBE-GGA even at higher plane wave cut-off energies.\cite{Lovvik} Thus the standard LDA/GGA functionals are unsuccessful in predicting the phase stability of KNO$_3$ polymorphs.\cite{Aydinol} Hence, we turn our attention to reproduce the experimental geometries thereby predicting the phase stability of the polymorphs using DFT-D2 method. In contrast to the above reports the obtained equilibrium volume using DFT-D2 method is in excellent agreement ($\sim$ 0.003-0.21 $\%$) with experimental results\cite{Nimmo,Adiwidjaja} and the calculated structural properties of KNO$_3$ polymorphs at various cut-off energies are presented in Table \ref{str2}. From Table \ref{str2}, we conclude that the dispersive interactions play a major role rather than number plane waves required for reproducing the experimental results for KNO$_3$ polymorphs. Also our total energy calculations show that $Pmcn$ and $Cmc2_1$ structures are differ by $\sim$ 1 meV/f.u. and this is consistent with the previous theoretical calculations and it is due to very similar local bonding environment (see figures \ref{crystal}f $\&$ g ) between the two structures.\cite{Lovvik} Since the two structures differ by a very small energy ($\sim$ 1 meV/f.u.), we have used $Pmcn$ unit cell for further calculations to reduce the computational effort. It is also found from our present calculations that the total energy of $\alpha$-phase is lowered by $\sim$ 7 meV/f.u. when compared $\gamma$-phase which shows the stability of $\alpha$-phase over $\gamma$-phase, which is in good agreement with experiments in contrast to previous calculations.\cite{Aydinol}

\begin{sidewaystable}
\centering
\caption{Calculated lattice constants (a, b, c, in $\AA$), lattice angle ($\alpha$, in $^o$) and volume (V, in $\AA^3$) of KNO$_3$ polymorphs with PBE-GGA and dispersion corrected (D2) functionals within the NCPP approach at various plane wave cut-off energies (E$_c$, in eV) along with experimental data. The relative errors (in $\%$) with respect to (w.r.t) experiments are given in parenthesis, here '-' and '+' signs indicate underestimation and overestimation of the calculated values w.r.t experiments.}
{\resizebox{0.99\textwidth}{!} {
\begin{tabular}{ccccccccccc}
 \hline   E$_c$  &           &    & \hspace{-0.7in}     750       &      & \hspace{-0.7in}  950         &     &  \hspace{-0.7in}  1250       &      &  \hspace{-0.7in}    1450   &             \\ \hline
 Phase                   & Parameter &   PBE    &  DFT-D2 &   PBE    &  DFT-D2  &   PBE    &  DFT-D2 &   PBE   &  DFT-D2   &  Expt.         \\ \hline
$Pmcn$                   &     a     &  5.488   &  5.453  &  5.483   &  5.451   &   5.483  &  5.452  &  5.485  &  5.455    &  5.4142$^a$    \\
($\alpha$-phase)         &           &  (+1.36)  & (+0.72)  & (+1.27)   & (+0.68)   &  (+1.27)  & (+0.70)  & (+1.31)  & (+0.75)    &        \\
                         &     b     &  9.256   &  9.246  &  9.272   &  9.256   &  9.263   &  9.252  &  9.268  &  9.252    &  9.1654$^a$    \\
                         &           &  (+0.99)  & (+0.88)  & (+1.16)   & (+0.99)   & (+1.06)   &  (+0.94) &  (+1.12) &  (+0.94)   &        \\
                         &     c     &  6.661   &  6.331  &  6.647   &  6.319   &  6.651   &  6.322  &  6.648  &  6.317    &  6.4309$^a$    \\
                         &           &  (+3.58)  &  (-1.55) &  (+3.36)  & (-1.74)   & (+3.42)   & (-1.69)  & (+3.38)  &  (-1.77)   &        \\
                         &     V     &  338.35  &  319.13 &  337.94  &  318.87  &  337.83  &  318.93 &  337.95 &  318.82   &  319.12$^a$    \\
                         &           &  (+6.03)  & (+0.003) &  (+5.90)  & (-0.08)   &  (+5.86)  & (-0.06)  & (+5.90)  &  (-0.09)   &        \\
$Cmc2_1$                 &     a     &   10.982 &  10.913 &  10.980  & 10.909   &   10.963  & 10.908  &  10.965 &  10.906   &  10.825$^b$   \\
(Super structure)        &           &   (+1.45) &  (+0.81) &  (+1.43)  & (+0.78)   &  (+1.27)  &  (+0.77) &  (+1.29) &  (+0.75)   &        \\
                         &     b     &  18.500  & 18.502  &  18.502  & 18.516   &  18.524   & 18.502  &  18.522 &  18.506   &  18.351$^b$   \\
                         &           &  (+0.81)  & (+0.82)  &  (+0.82)  & (+0.90)   &    (+0.94)    & (+0.82)  & (+0.93)  &  (+0.89)   &    \\
                         &     c     &  6.658   &  6.321  &  6.658   & 6.316    &  6.657  & 6.321   &  6.658  &  6.322    &  6.435$^b$      \\
                         &           &  (+3.46)  &  (-1.77) &  (+3.46)  & (-1.85)   &   (+3.45)     & (-1.77)  &  (+3.46) &  (-1.76)   &    \\
                         &     V     &  1352.79 & 1276.22 &  1352.55 &  1275.76 &  1351.83  & 1275.59 & 1352.32 &  1275.95  &  1278.31$^b$  \\
                         &           &   (+5.83) &  (-0.16) &  (+5.81)  & (-0.20)   &    (+5.75)    &  (-0.21) &  (+5.79) &  (-0.18)   &    \\
$R3m$                    &     a     &  4.492   &  4.317  &  4.488   &  4.310   &  4.487   &  4.314  &  4.489  &  4.310    &  4.3699$^c$    \\
($\gamma$-phase)         &           &  (+2.79)  & (-1.21)  &  (+2.70)  &  (-1.37)  &  (+2.68)  &  (-1.28) &  (+2.72) &  (-1.37)   &        \\
                         & $\alpha$  &  75.71   &  79.32  &  75.73   &  79.58   &  75.76   &  79.40  &  75.71  &  79.54    &  76.56$^c$     \\
                         &           &  (-1.11)  & (+3.60)  &  (-1.08)  &  (+3.94)  &  (-1.04)  &  (+3.71) &  (-1.11) &  (+3.89)   &        \\ \hline
\end{tabular}}}
\\ $^a$Ref.\cite{Nimmo},
$^b$Ref.\cite{Adiwidjaja},
$^c$Ref.\cite{Nimmo3}
\label{str2}
\end{sidewaystable}

\subsection{Elastic constants and Mechanical properties}
It is of great interest to determine the relation between elastic and other physical and energetic properties of crystals.\cite{Belomestnykh} Among the physical constants, elastic constants are fundamental parameters for crystalline solids which describe stiffness of the solid against externally applied strains and also provide information about the mechanical and dynamical behavior of crystals as well as the nature of forces operating in solids. The elastic constants are calculated using volume-conserving strain technique\cite{Mehl} at the DFT-D2 equilibrium volume for all the examined compounds. A completely asymmetric crystal can be described by 21 independent elastic constants. Due to the rhombohedral (LiNO$_3$ and NaNO$_3$), orthorhombic (KNO$_3$), cubic (NaClO$_3$) and monoclinic (KClO$_3$) symmetry of crystals, they possess 6, 9, 3 and 13 independent elastic constants, respectively. The obtained elastic constants are in good agreement with Ultrasonic Pulse-Echo measurements\cite{Haussuhl,Michard,Viswanathan} and are presented in Table \ref{elastic}. The computed elastic moduli are comparable with the ones obtained from inter-atomic potentials by including non-bonded interactions in the calculations\cite{Mort} which shows the importance of dispersive interactions in these class of materials. The elastic constants for alkali metal nitrates (LiNO$_3$, NaNO$_3$ and KNO$_3$) are closely comparable (in magnitude) with corresponding alkali metal azides (LiN$_3$, NaN$_3$ and KN$_3$)\cite{ramesh1,ramesh3} and for alkali metal chlorates (NaClO$_3$ and KClO$_3$) with corresponding alkali halides (NaCl and KCl)\cite{Srivastava}. Direct comparison is not possible due to distinct crystal symmetry of the studied systems with the corresponding alkali metal azides and halides.

\begin{table}
\centering
\caption{Calculated single crystal elastic constants (C$_{ij}$, in GPa) and polycrystalline bulk moduli (B, in GPa), density ($\rho$, in g/cc), wave velocities ($v_{l}$, $v_{t}$ and $v_{m}$, in km/s) and Debye temperature ($\theta_{D}$, in K) of LiNO$_3$, NaNO$_3$, KNO$_3$, NaClO$_3$, KClO$_3$ at the DFT-D2 equilibrium volume. The experimental values are given in parenthesis for C$_{ij}$, $\rho$ and other calculations for $\theta_{D}$}
{\resizebox{0.99\textwidth}{!}{
\begin{tabular}{cccccc}
\hline Parameter &  LiNO$_3$$^a$    &  NaNO$_3$$^a$  &  KNO$_3$$^{a,b}$     &   NaClO$_3$$^c$ &  KClO$_3$  \\ \hline
C$_{11}$         &  112.50 (97.06)  & 62.30 (56.81)  & 45.50 (37.16, 35.8)  &   57.75 (49.2)  &  30.82     \\
C$_{22}$         &       -          &       -        & 37.36 (29.89, 30.0)  &       -         &  46.69     \\
C$_{33}$         &  69.35 (66.59)   & 27.68 (32.84)  & 23.33 (20.37, 20.4)  &       -         &  41.24     \\
C$_{44}$         &  24.31 (16.75)   & 14.35 (11.97)  & 9.71  (6.80, 6.7)    &   10.21 (11.6)  &  9.97      \\
C$_{55}$         &      -           &      -         & 8.25  (5.43, 5.4)    &       -         &  8.66      \\
C$_{66}$         &  41.00 (33.78)   & 21.42 (17.90)  & 8.43  (8.35, 8.3)    &       -         &  17.38     \\
C$_{12}$         &  30.49 (30.09)   & 19.45 (21.01)  & 23.00 (16.80, 13.4)  &   18.23 (14.2)  &  20.67     \\
C$_{13}$         &  23.18 (14.96)   & 17.48 (17.96)  & 11.33 (10.96, 11.6)  &        -        &  11.36     \\
C$_{23}$         &      -           &      -         & 13.27 (11.14, 9.2)   &        -        &  9.57      \\
C$_{14}$         &  7.96 (-2.75)    & 9.43  (-7.89)  &        13.29         &        8.02     &  10.7      \\
B                &  48.28           &    26.80       &  20.88               &       31.40     &  20.84     \\
$\rho$           & 2.313 (2.368)$^d$ & 2.200 (2.261)$^e$ &   2.119 (2.104)$^f$ &  2.463 (2.486)$^g$ &  2.397 (2.338) $^h$    \\
$v_{l}$          &   6.21           &   4.55         &   3.94               &     4.47        &   3.85     \\
$v_{t}$          &   3.64           &   2.52         &   2.06               &     2.33        &   2.15     \\
$v_{m}$          &   4.04           &   2.81         &   2.30               &     2.60        &   2.39     \\
$\theta_{D}$(K)  &   559(492)$^i$   &   358(368)$^i$ &   273(256)$^i$       &  319            &  277   \\ \hline
\end{tabular}}}
\\ $^a$Ref.\cite{Haussuhl},
$^b$Ref.\cite{Michard},
$^c$Ref.\cite{Viswanathan},
$^d$Ref.\cite{Wu},
$^e$Ref.\cite{Gonschorek},
$^f$Ref.\cite{Nimmo},
$^g$Ref.\cite{Abrahams},
$^h$Ref.\cite{Danielsen},
$^i$Ref.\cite{Belomestnykh}
\label{elastic}
\end{table}

The well-known Born stability criteria are a set of conditions on the elastic constants (C$_{ij}$) to verify the mechanical stability of a crystal which are related to the second-order change in the internal energy of a crystal under deformation. All the investigated crystal systems obey Born's mechanical stability criteria\cite{Born} indicating that these materials are mechanically stable at ambient conditions. The elastic constants are primarily classified into longitudinal (C$_{11}$, C$_{22}$, C$_{33}$), transverse (C$_{12}$, C$_{13}$, C$_{23}$) and shear (C$_{44}$, C$_{55}$, C$_{66}$) elastic constants. The longitudinal elastic constants are found to follow the order C$_{33}$ $\textless$ C$_{11}$ (C$_{33}$ $\textless$ C$_{22}$ $\textless$ C$_{11}$ for KNO$_3$), which implies that the crystals exhibit anisotropic compressibility along various crystallographic directions $i.e.$ weak intermolecular interactions along c-axis over a-axis is observed for nitrates and this is in very good agreement with the experimental observation\cite{Haussuhl}. The alkali metal nitrates such as LiNO$_3$, NaNO$_3$ and KNO$_3$ consist of alternate layers of metal atom and NO$_3$ group stacked along c-axis and strong covalent bonding within NO$_3$ group which is oriented parallel to (001). While NaClO$_3$ possesses isotropic elastic constants due to its cubic crystal symmetry whereas KClO$_3$ shows strong anisotropy C$_{11}$ $\textless$ C$_{33}$ $\textless$ C$_{22}$ along three crystallographic directions which shows the anisotropic compressibility of the investigated systems. The derived bulk moduli from elastic moduli are comparable with the corresponding alkali metal halides\cite{Haussuhl1} 36.9, 28.6, and 20.8 GPa for LiCl, NaCl and KCl, respectively. The bulk moduli of nitrate/chlorate ion with light alkali metal atom show large deviation but it is closely comparable with the corresponding alkali halide on moving from Li to K. The calculated elastic and bulk moduli for the investigated compounds decrease in the following order; LiNO$_3$ $\textgreater$ NaNO$_3$ $\textgreater$ KNO$_3$ for nitrates and NaClO$_3$ $\textgreater$ KClO$_3$ for chlorates, which is due to the increase in ionic size of the metal atom. In addition, we also estimated the Debye temperature ($\Theta_D$) which is a fundamental quantity that correlates many physical properties such as specific heat and melting point of solids from the calculated elastic constants data. It also determines thermal characteristics of the material, a high value of $\Theta_D$ implies higher thermal conductivity. At low temperature, $\Theta_D$ can be estimated from the longitudinal, transverse and average wave velocities as given in Table \ref{elastic}. The obtained  $\Theta_D$ values decrease from Li $\rightarrow$ Na $\rightarrow$ K (Na $\rightarrow$ K) for nitrates (chlorates) and are consistent with previous calculations on alkali metal nitrates,\cite{Belomestnykh} and a high value of LiNO$_3$ shows its higher thermal conductivity than other studied compounds.

\subsubsection{Zone center phonon frequencies, IR and Raman spectra}
\begin{figure}
\centering
\includegraphics[height = 5.0in, width=6.0in]{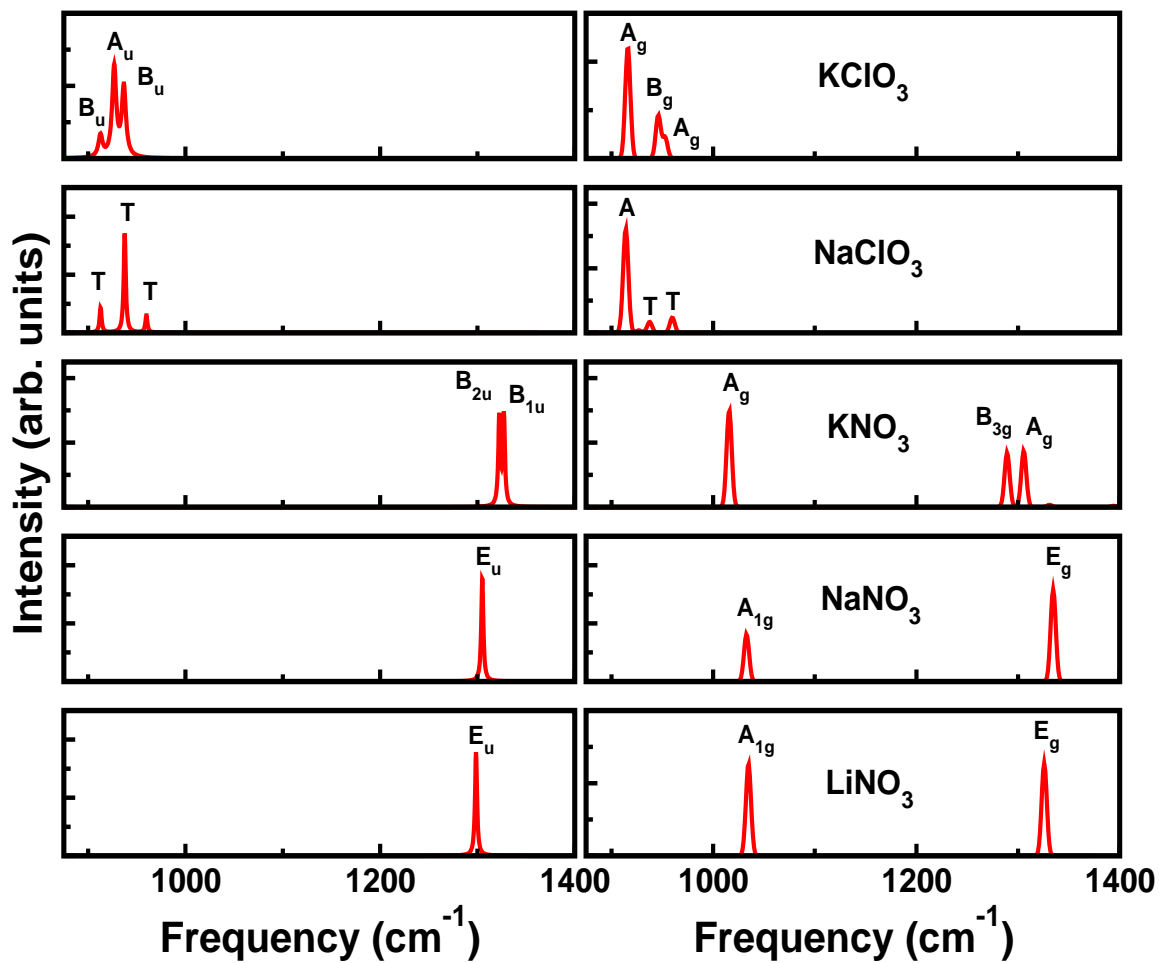}
\caption{\label{internal} Calculated zone centre IR (left) and Raman (right) spectra of internal modes of LiNO$_3$, NaNO$_3$, KNO$_3$, NaClO$_3$, and KClO$_3$.}
\label{internal}
\end{figure}

In our previous study,\cite{kondal} we have used LDA (CA-PZ), GGA (PBE) and DFT-D2 functionals to calculate the vibrational frequencies of KClO$_3$. The obtained lattice mode frequencies using CA-PZ and PBE functionals show large deviation from the experimental results as the functionals are inadequate to treat intermolecular interactions in KClO$_3$. Hence, in the present study, we made a detailed analysis of vibrational spectra using vibrational frequencies obtained from DFT-D2 method for all these compounds using the linear response approach within density functional perturbation theory. MNO$_3$ (M = Li, Na, K) and MClO$_3$ (M = Na, K) compounds consist of two anions namely nitrate (NO$_3$) and chlorate (ClO$_3$). NO$_3$ anion consists of four internal vibrations (are in cm$^{-1}$) in its free state such as Raman active 1050 (A$_1$), IR active 820 (A$_2$), and both Raman and IR active 1385 (E), 725 (E) degenerate modes. Also ClO$_3$ ion possesses four internal vibrations (are in cm$^{-1}$) such as 930 (A$_1$), 615 (A$_2$), 975 (E), and 478 (E) similar to nitrate ion in its unbound state and these internal modes are both Raman and IR active.\cite{Ramdas1} The internal degrees of freedom increase when transformation from free molecular (NO$_3$/ClO$_3$ in the present case) ion to a crystal due to increase in number of molecules per unit cell. The unit cells of LiNO$_3$ and NaNO$_3$, KNO$_3$, NaClO$_3$, and KClO$_3$ contain 10, 20, 20 and 10 (N) atoms respectively giving rise to a total of 30, 60, 60, 30 (3  acoustic and 3N-3 optical) vibrational modes. Further optical modes are classified into internal ($\textgreater$ 400 cm$^{-1}$) and external (lattice) ($\textless$ 400 cm$^{-1}$) modes, where these internal modes arise from anions and lattice modes are from the whole crystal lattice. According to group theory analysis of the respective space groups, the group symmetry decomposition into irreducible representations of the modes for all the investigated compounds are given in Table \ref{rep}. For LiNO$_3$ and NaNO$_3$, A$_{1g}$ and E$_{g}$ are Raman active, A$_{2u}$ and E$_{u}$ are IR active and the A$_{2g}$ and A$_{1u}$ are forbidden in both. The transformation of free molecular NO$_3$ anion to (Li/Na)NO$_3$ crystal, which consists of two molecules within unit cell leads to an increase from 4 $\rightarrow$ 8 distinct internal modes in its crystalline form, in which E$_g$ and E$_u$ are doubly degenerate modes and the detailed factor group analysis given in Ref.\cite{Khanna} by Khanna. While there are ten different lattice modes with three doubly degenerate modes (A$_{2u}$, A$_{2g}$, E$_g$) and the calculated lattice modes are compared with the available experimental data\cite{Nakagawa} which are given in Table 1 of the supplementary material. For KNO$_3$, A$_{g}$, B$_{1g}$, B$_{2g}$ are Raman active, B$_{1u}$, B$_{2u}$, B$_{3u}$ are IR active and A$_{u}$ is forbidden in both. The presence of NO$_3$ anion with K$^+$ ion in KNO$_3$ crystal removes the degeneracy of NO$_3$ ion internal vibrations from 4 $\rightarrow$ 24 non-degenerate modes due to four molecules in the KNO$_3$ crystal as presented in Table 5. We also made lattice vibrational mode  assignment (33 in number) and compared them with the available experimental data\cite{Liu} as presented in Table 2 of the supplementary material. In case of alkali metal chlorates, A, E are Raman active and T is active in both for NaClO$_3$, whereas A$_g$, B$_g$ are Raman active and A$_u$, B$_u$ are IR active for KClO$_3$. Existence of ClO$_3$ ion along with Na(K) ion in NaClO$_3$ (KClO$_3$) crystal increases the internal vibrations from 4 $\rightarrow$ 10 (12) non equivalent internal vibrations. Among 10 internal modes of NaClO$_3$, T and E are triply and doubly degenerate modes respectively whereas all the 12 modes are non-degenerate in case of KClO$_3$. The remaining 33 optical lattice modes of NaClO$_3$ and their vibrational assignments are given in Table 3 of the supplementary material along with the available experimental data.\cite{Hartwig} A detailed lattice vibrational analysis for KClO$_3$ is given in our previous study.\cite{kondal} We have also calculated IR and Raman spectra for the investigated compounds, the internal and lattice modes are given in figures \ref{internal} and \ref{lattice}, respectively. Also it can be noticed that the internal mode frequencies do not differ greatly when comparing within the nitrates (LiNO$_3$, NaNO$_3$, KNO$_3$) and chlorates (NaClO$_3$, KClO$_3$) which shows the free nature of the nitrate/chlorate ion in the highly ionic salts and the internal modes are purely dominated by nitrate/chlorate group. While replacement of nitrogen by chlorine atom $i.e.$ moving from nitrates (NaNO$_3$, KNO$_3$) $\rightarrow$ chlorates (NaClO$_3$, KClO$_3$), we observe a huge reduction in the symmetric and asymmetric stretching internal vibrations due to presence of heavy non-metal chlorine atom. The presence of metal atom affect only the lattice mode frequencies, which decreases with increase in mass (Li $\textgreater$ Na $\textgreater$ K) of the metal atom in both nitrates and chlorates and this can be clearly seen from figure \ref{lattice}.

\begin{table}
\centering
\caption{The group symmetry decomposition into irreducible representations of the vibrational modes for LiNO$_3$, NaNO$_3$, KNO$_3$, NaClO$_3$, KClO$_3$}
\begin{tabular}{cc}
\hline Compound      &    Irr. Rep.   \\ \hline
(Li/Na)NO$_3$      &  $\Gamma_{acc}$ = A$_{2u}$ $\oplus$ 2E$_u$   \\
(R$\bar{3}$c, Z=6) &  $\Gamma_{optic}$ = 8E$_g$ $\oplus$ 10E$_u$ $\oplus$ 3A$_{2u}$ $\oplus$ 3A$_{2g}$ $\oplus$ 2A$_{1u}$ $\oplus$ A$_{1g}$ \\
KNO$_3$            &  $\Gamma_{acc}$ = B$_{1u}$ $\oplus$ B$_{2u}$ $\oplus$ B$_{3u}$    \\
(Pnma, Z=4)        & $\Gamma_{optic}$ = 8B$_{1u}$ $\oplus$ 5B$_{2u}$ $\oplus$ 8B$_{3u}$ $\oplus$ 6B$_{1g}$ $\oplus$ 9B$_{2g}$ $\oplus$ 6B$_{3g}$ $\oplus$ 9A$_g$ $\oplus$ 6A$_u$ \\
NaClO$_3$          &  $\Gamma_{acc}$ = 3T   \\
(P2$_1$3, Z=4)     &  $\Gamma_{optic}$ = 42T $\oplus$ 10E $\oplus$ 5A \\
KClO$_3$           &  $\Gamma_{acc}$ = A$_u$ $\oplus$ 2B$_u$ \\
(P2$_1$/m, Z=2)    &  $\Gamma_{optic}$ = 9A$_g$ $\oplus$ 5A$_u$ $\oplus$ 6B$_g$ $\oplus$ 7B$_u$ \\ \hline
\end{tabular}
\label{rep}
\end{table}

\begin{figure}
\centering
\includegraphics[height = 5.0in, width=6.0in]{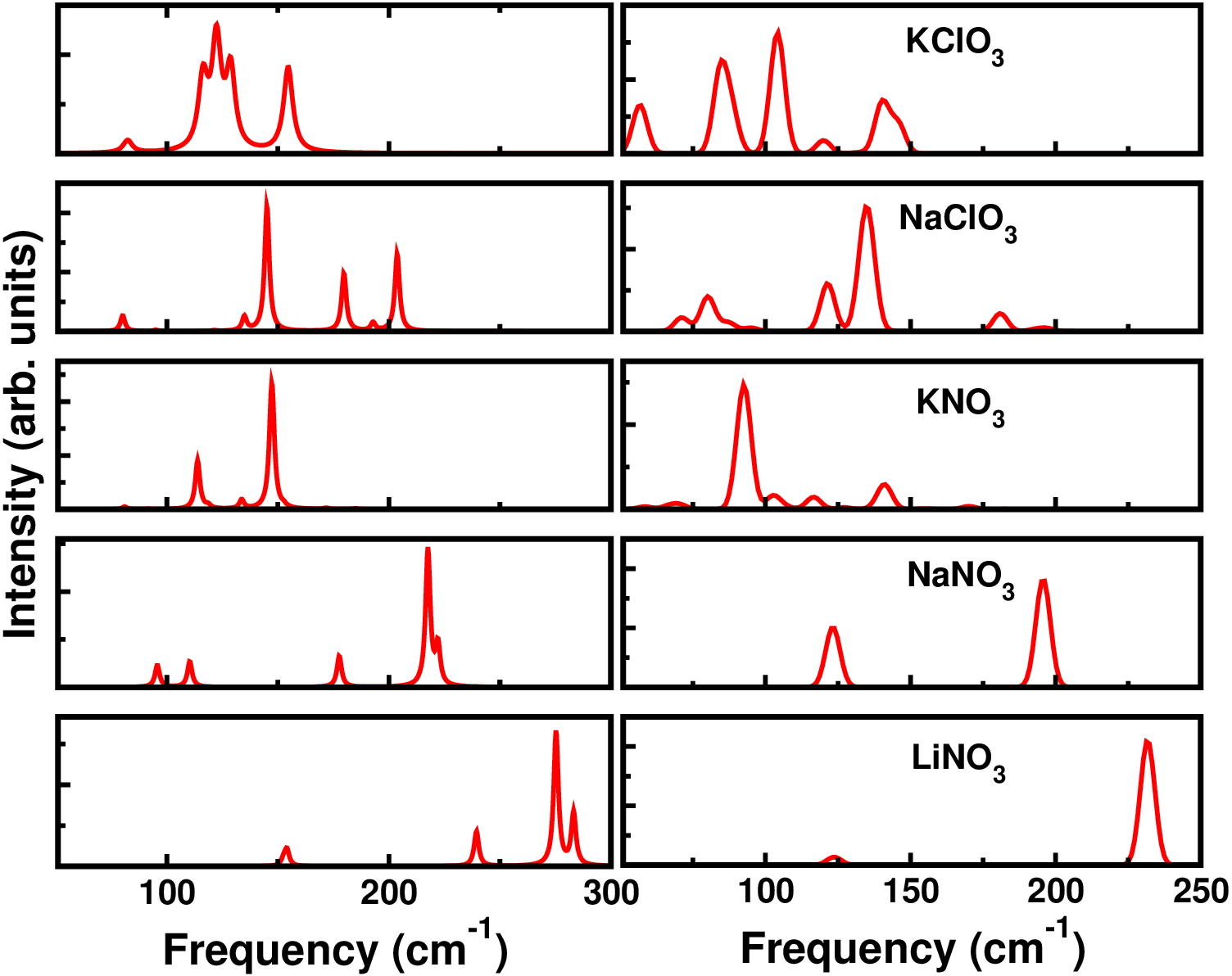}
\caption{\label{lattice} Calculated zone centre IR (left) and Raman (right) spectra of lattice modes of LiNO$_3$, NaNO$_3$, KNO$_3$, NaClO$_3$, and KClO$_3$.}
\label{lattice}
\end{figure}

\begin{sidewaystable}
\caption{Calculated asymmetric ($\nu_{Asym}$), symmetric ($\nu_{Sym}$) stretching and bending ($\nu_{Bend}$) vibrational frequencies (in cm$^{-1}$) of LiNO$_3$, NaNO$_3$, KNO$_3$ and NaClO$_3$, KClO$_3$ compounds at the DFT-D2 equilibrium volume}
{\resizebox{0.80\textwidth}{!} {\begin{minipage}{\textwidth}
\centering
\begin{tabular}{ccccccccccc}
\hline  & \hspace{-0.8in} LiNO$_3$  &         & \hspace{-0.8in} NaNO$_3$ &        &  \hspace{-0.8in} KNO$_3$ &      &  \hspace{-0.8in} NaClO$_3$ &     & \hspace{-0.8in} KClO$_3$ &     \\ \hline
 IrrRep &  Freq   &  IrrRep  &  Freq   &  IrrRep   & Freq   & IrrRep & Freq &  IrrRep & Freq   &   Assignment                         \\ \hline
 E$_g$  & 1325.8  & E$_g$    & 1334.6  &  B$_{2g}$ & 1391.8 &  T   & 959.8  &  B$_g$  & 952.8  &   $\nu_{Asym}$                       \\
  & (1375$^a$,1386$^b$)  &     & (1385$^{b,c}$,1386$^b$)    &          &        &      &        &    &   (982$^{g,h}$,983$^i$) &      \\
 E$_u$  & 1298.6  & E$_u$    & 1305.0  &  B$_{1g}$ & 1335.9 &  T   & 937.6  &  B$_u$  & 936.7  &                                      \\
        & (1380$^a$)  & & (1405$^a$,1365$^b$) & &  (1348$^d$) &     &        &         & (1000$^g$) &                                 \\
        &         &          &         &  B$_{2u}$ & 1327.3 &  E   & 926.9  &  A$_g$  & 946.0  &                                      \\
        &         &          &         &           &        &      & (958$^e$,957$^f$)  &     & (978$^h$,979$^{g,i}$)  &              \\
        &         &          &         &  B$_{1u}$ & 1324.9 &      &        &  A$_u$  & 926.7  &                                      \\
        &         &          &         &    A$_g$  & 1302.9 &      &        &         &  (992$^g$)  &                                 \\
        &         &          &         &     & (1360$^d$) &        &        &         &        &                                      \\
        &         &          &         &  B$_{3u}$ & 1298.3 &      &        &         &        &                                      \\
        &         &          &         &  B$_{3g}$ & 1294.2 &      &        &         &        &                                      \\
        &         &          &         &   A$_u$   & 1282.9 &      &        &         &        &                                      \\
A$_{1g}$ & 1034.8 & A$_{1g}$ & 1032.9  &   A$_g$   & 1015.2 &  A   & 914.3  &  A$_g$  & 916.1  &    $\nu_{Sym}$                       \\
& (1067$^a$,1073$^b$) & & (1068$^{a,c}$,1069$^b$)&&(1050$^b$,1054$^d$)&& (931.5$^e$, 937$^f$) &    & (939$^h$,940$^{g,i}$) &          \\
A$_{1u}$ & 1033.4 & A$_{1u}$ & 1032.0  &  B$_{3u}$ & 1014.9 &  T   & 912.8  &  B$_u$  & 912.6  &                                      \\
         &        &          &         &  B$_{2g}$ & 1013.8 &      &        &         & (939$^f$)  &                                  \\
         &        &          &         &  B$_{1u}$ & 1013.6 &                                                                         \\
A$_{2g}$ & 800.0  & A$_{2g}$ & 800.5   &  B$_{1u}$ & 812.3  &  T   & 600.4  &  A$_g$  &  595.3 &   $\nu_{Bend}$                       \\
         &        &          &         &           &        &      &        &         & (619$^h$,620$^{g,i}$)     &                   \\
A$_{2u}$ & 798.5  & A$_{2u}$ & 799.2   &  B$_{2g}$ & 811.9  &  A   & 597.1  &  B$_u$  &  593.7 &                                      \\
      & (837$^{a,b}$)  &          & (831$^a$) &    &        &      & (620$^e$,618$^f$)  &     &  (620$^g$)   &                        \\
E$_{u}$  & 710.5  & E$_{u}$  & 702.7   &  A$_g$    & 793.9  &  T   & 462.6  &  B$_u$  &  467.3 &                                      \\
     & (730$^a$,732$^b$)  &      & (692$^a$,723$^b$)   &   & (827$^d$)  &      &        &  &  (493$^g$) &                             \\
E$_{g}$  & 708.0  & E$_{g}$  & 701.4   &  B$_{3u}$ & 793.4  &  T   & 459.2  &  A$_g$  &  463.6 &                                      \\
     & (726$^a$,737$^b$)  &      & (724$^{a,c}$,728$^b$)   &       &        &      &  &     & (488$^g$,490$^i$)     &                 \\
         &        &          &         &  B$_{3u}$ & 691.6  &  E   & 456.5  &  B$_g$  &  463.8 &                                      \\
         &        &          &         &           &        &      & (481.5$^e$,482$^f$)&     &  (487$^g$) &                          \\
         &        &          &         &  B$_{2u}$ & 690.9  &      &        &  A$_u$  &  463.7 &                                      \\
         &        &          &         &  B$_{1g}$ & 690.7  &      &        &         &  (484$^g$) &                                  \\
         &        &          &         &  A$_{u}$  & 690.6  &                                                                         \\
         &        &          &         &  B$_{2g}$ & 690.6  &                                                                         \\
         &        &          &         &  B$_{3g}$ & 690.2  &                                                                         \\
         &        &          &         &  B$_{1u}$ & 690.1  &                                                                         \\
         &        &          &         &  A$_{g}$  & 689.2  &                                                                         \\
         &        &          &         &           &(718$^d$) &                                                                       \\ \hline
\end{tabular}
\\ $^a$Ref.\cite{Nakagawa},
$^b$Ref.\cite{James},
$^c$Ref.\cite{Rousseau},
$^d$Ref.\cite{Liu},
$^e$Ref.\cite{Brooker1},
$^f$Ref.\cite{Hartwig}
$^g$Ref.\cite{Bates},
$^h$Ref.\cite{Brooker2},
$^i$Ref.\cite{Heyns}
\end{minipage}}}
\label{int}
\end{sidewaystable}

\par Overall, we noticed that the presence of nitrate ion in MNO$_3$ (M = Li, Na, K) and chlorate ion in MClO$_3$ (M = Na, K) crystals increases the number of internal degrees (8, 24 and 10, 12) of freedom when compared to their free state depending upon the number of molecules per unit cell in the respective crystal structures. Irrespective of distinct internal vibrations which arise form various compounds, the internal modes are classified into three branches such as asymmetric, symmetric stretching and bending vibrations for all the studied compounds. The calculated internal (Table 5 and figure \ref{Vib}) and lattice (Tables 1, 2 and 3 of the supplementary material) vibrations and their assignment for all these compounds are in good agreement with the available experimental data.\cite{Nakagawa,James,Rousseau,Liu,Brooker1,Hartwig,Bates,Brooker2,Heyns}

\begin{figure}
\centering
\includegraphics[height = 8in,width=6in]{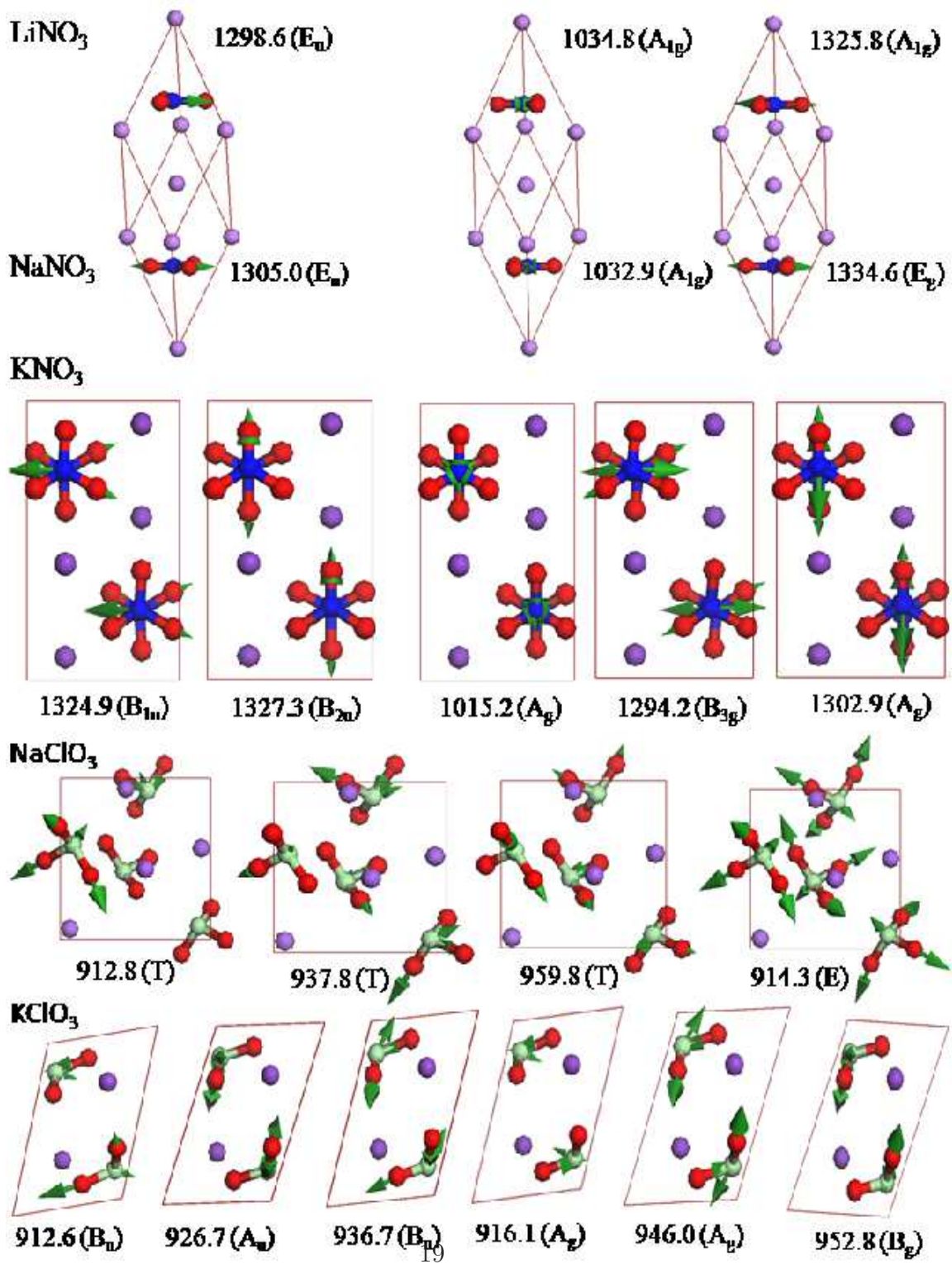}
\caption{\label{Vib} Vibrational animations of internal IR and Raman active modes for the solid oxidizers.}
\label{Vib}
\end{figure}

\subsubsection{Electronic structure and chemical bonding}
\begin{figure}
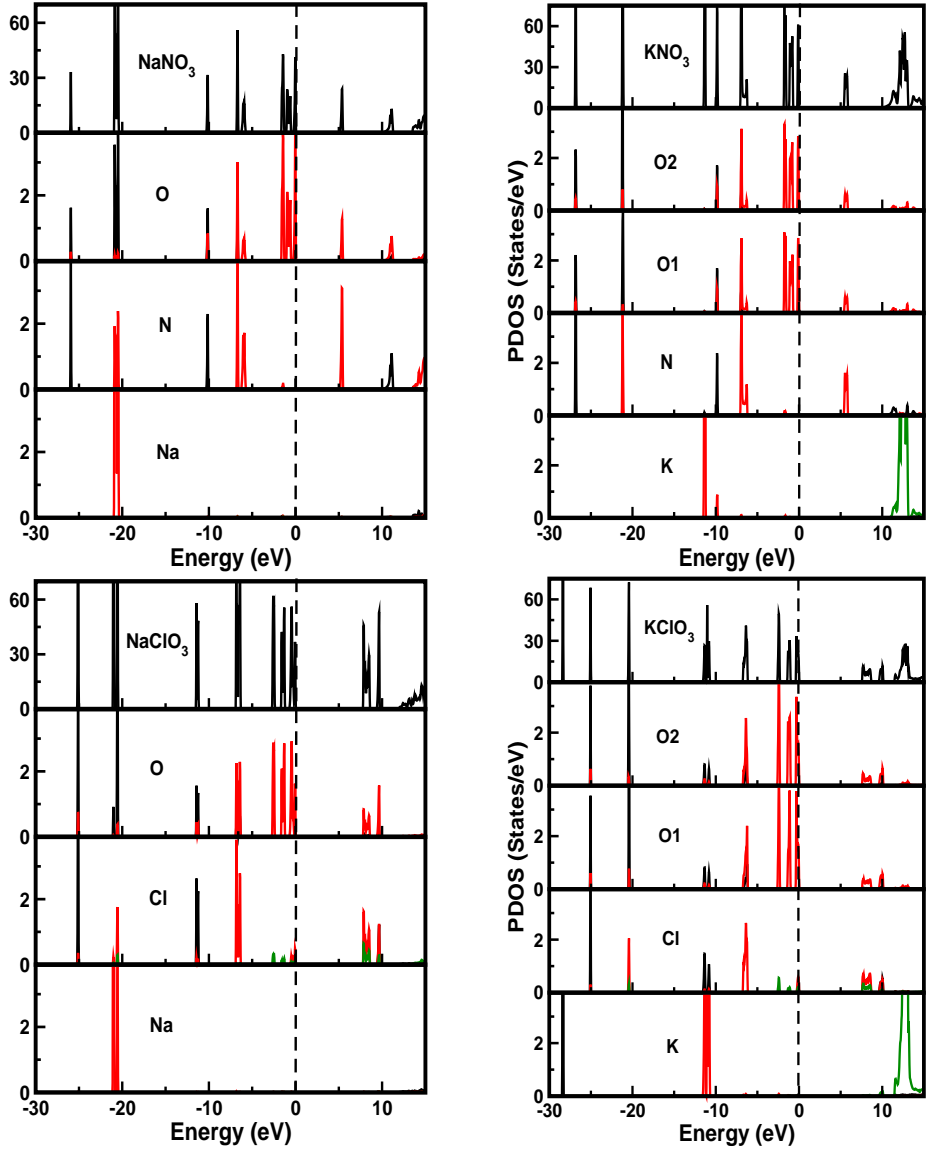

\centering
\includegraphics[height = 3.0in, width=2.2in]{NaNO3_DOS.eps}  \hspace{0.3in}
\includegraphics[height = 3.0in, width=2.2in]{KNO3_DOS.eps} \\
\includegraphics[height = 3.0in, width=2.2in]{NaClO3_DOS.eps}  \hspace{0.3 in}
\includegraphics[height = 3.0in, width=2.2in]{KClO3_DOS.eps}
\caption{\label{DOS} Calculated electronic total and partial density of states of (a) NaNO$_3$, (b) KNO$_3$ (c) NaClO$_3$ and (d) KClO$_3$ using TB-mBJ potential at the experimental crystal structures. }
\label{DOS}
\end{figure}
Electronic structure calculations not only predict physical and chemical properties of materials but also provide a good insight to the experimentalists to synthesize novel materials for various applications. Extensive theoretical studies are devoted to understand the electronic structure of bulk and surfaces of oxianionic alkali metal nitrates, chlorates and perchlorates.\cite{Zhuravlev1,Zhuravlev2,Zhuravlev3,Zhuravlev4,Zhuravlev5,Lovvik,Pradeep} However, the reported band gaps for these materials are severely under- and over-estimated  using standard LDA/GGA functionals and Hatree-Fock methods, respectively. Therefore, we have used TB-mBJ potential to get reliable energy bands for the investigated materials.
\par The electronic band structure of MNO$_3$ (M = Li, Na, and K) and MClO$_3$ (M = Na, K) compounds at the experimental lattice constants is calculated using three functionals such as PBE, EV-GGA, and TB-mBJ. The calculated TB-mBJ band structures show that the investigated oxyanionic crystals are wide band gap insulators. The band gaps are reported in Table \ref{gaps} using three functionals, from this Table we find that TB-mBJ potential improves the band gaps when compared to PBE and EV-GGA functionals and are comparable with the available experimental data\cite{Vorobeva,Guravlev} with an exception for NaNO$_3$ in which we found relatively large discrepancy between predicted band gap and experimental values. However, PBE-GGA band gaps are consistent with previously reported\cite{Zhuravlev1,Aydinol,Zhuravlev2,Lovvik,Pradeep,Zhuravlev4} values for these materials. It is also found that (Li/Na)NO$_3$ show indirect gap along F-$\Gamma$ direction whereas KNO$_3$, NaClO$_3$ and KClO$_3$ are direct band gap insulators along $\Gamma$, X, and A directions, respectively. Also, the nature of chemical bonding in these materials is explained from calculated partial density of states (PDOS) as shown in figure \ref{DOS}. The top of the valence band (near Fermi level) is completely dominated by O-2p and less contribution from N-p(Cl-p and d) states in alkali metal nitrates (chlorates). It can be clearly seen from figure \ref{DOS} that there is a strong overlap between p-states of N/Cl and O atoms around -7 eV, s-states of N/Cl and O atoms around -11.5 eV, p-states of N/Cl and s-states of O-atoms around -21 eV and s-states of N/Cl and O atoms about -25.5 eV indicating that there exits a strong covalent bond between N-O and Cl-O within the nitrate and chlorate group, respectively. We also noticed that s-states are derived from K atom only in KClO$_3$ about -28 eV. In the valence band region, p-states of Na and K atoms are located around -20/-10 eV for Na/K based compounds which implies that the metal states are separated from the nitrate/chlorate group which indicates an ionic bonding between metal cation and nitrate/chlorate anion. Conduction band is mainly dominated by d-states of Cl and K, and s,p-states of N/Cl and O atoms. The bands are very narrow (flat) as commonly expected for ionic crystals. Overall, we observe mixed bonding nature $i.e.$ predominant ionic bonding between metal (Li, Na, K) cation and NO$_3$/ClO$_3$ anion and covalent bonding between N/Cl and O atoms within the NO$_3$/ClO$_3$ anionic group in the bulk oxyanionic crystals which is consistent with the previous theoretical calculations on these materials.\cite{Zhuravlev1,Aydinol,Zhuravlev2,Lovvik,Pradeep,Zhuravlev4}

\begin{table}
\centering
\caption{Calculated electronic band gaps (in eV) of LiNO$_3$, NaNO$_3$, KNO$_3$, NaClO$_3$, KClO$_3$ using three (PBE-GGA, EV-GGA and TB-mBJ) functionals}
{\resizebox{0.99\textwidth}{!}{
\begin{tabular}{cccccc}
\hline Functional  & LiNO$_3$  &   NaNO$_3$  &  KNO$_3$  &  NaClO$_3$  &  KClO$_3$    \\ \hline
 PBE-GGA    &  2.98     &    2.99     &  3.09     &   5.89      &   5.67       \\
 EV-GGA     &  3.16     &    3.17     &  3.26     &   6.11      &   5.89       \\
 TB-mBJ     &  4.94     &    5.16     &  5.38     &   7.84      &   7.58       \\
 Others     &  2.4$^a$  &    1.4$^a$  &  1.68$^b$, 2.6$^c$, 3.0$^d$, 3.07$^{e,f}$  & 3.0$^c$ & 5.66$^f$ \\
            &           &             &  3.24$^e$, 5.40$^e$, 14.98$^e$ &  &    \\                         
 Expt.      &   -       &   $\sim$6-7$^g$, 8.2$^h$  &  $\sim$6-7$^g$ &   -          &  8$^h$       \\ \hline
\end{tabular}}}
\\ $^a$Ref.\cite{Zhuravlev1},
$^b$Ref.\cite{Aydinol},
$^c$Ref.\cite{Zhuravlev2},
$^d$Ref.\cite{Lovvik},
$^e$Ref.\cite{Pradeep},
$^f$Ref.\cite{Zhuravlev4},
$^g$Ref.\cite{Guravlev},
$^h$Ref.\cite{Vorobeva}
\label{gaps}
\end{table}

\subsubsection{Optical properties}

\begin{figure}
\centering
\includegraphics[height = 4.5in,width=5.5in]{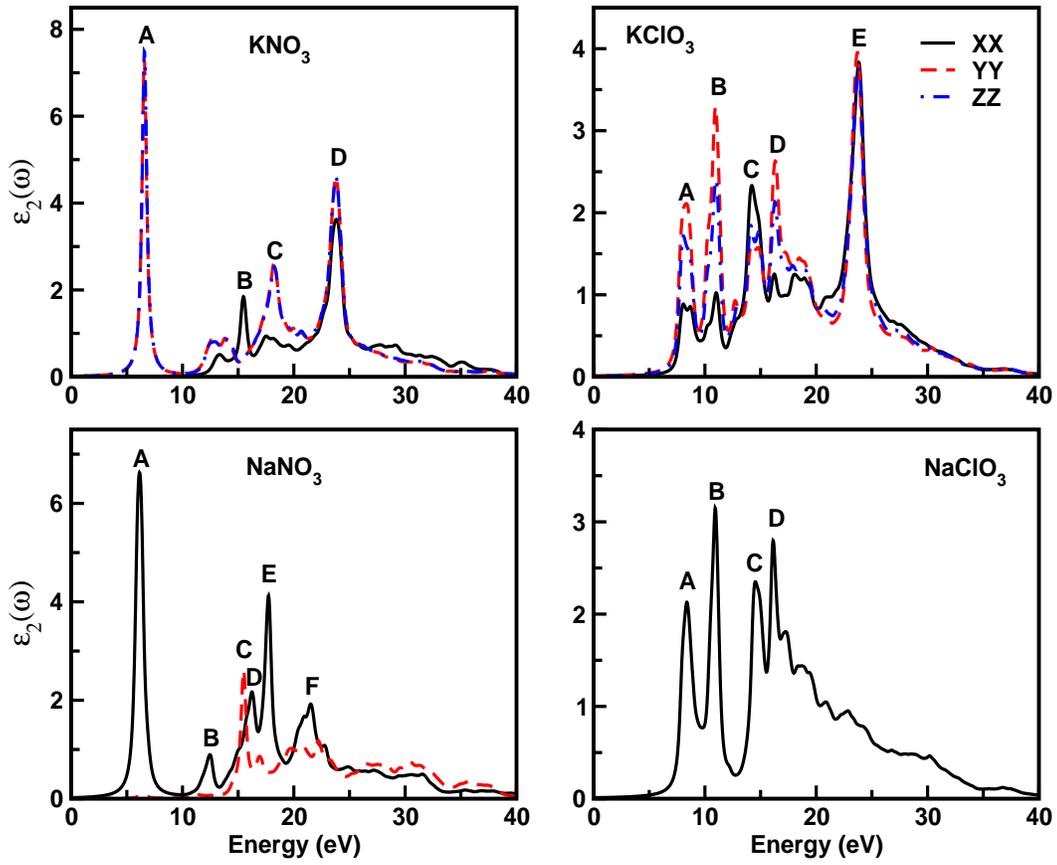}
\caption{\label{Epsilon} Calculated imaginary part of dielectric function for NaNO$_3$, KNO$_3$, NaClO$_3$, and KClO$_3$ using TB-mBJ potential at the experimental crystal structures. }
\label{Epsilon}
\end{figure}

In order to understand photo sensitive decomposition mechanisms in the energetic formulations, we try to shed more light on optical properties of these materials. Very limited studies are available in the literature to explore the optical properties of these oxidizers. Vorobeva et al\cite{Vorobeva} measured the reflection spectra and quantum yield photo emission of alkali metal nitrates and chlorates and they reported that anionic and cationic excitations occur in the low ($\textless$ 20 eV) and high ($\textgreater$ 20 eV) energy region, respectively. Photo emission and reflection spectra of MNO$_3$ (M = Na, K, Rb, and Cs)\cite{Zhuravlvev} and NaClO$_3$\cite{Burkov} were measured in the fundamental absorption region. Optical spectra of metal (Li, Na, K) nitrites, nitrates, chlorates, perchlorates, sulfites, and sulfates were calculated using CRYSTAL package within LDA. As discussed in the above section the band gaps are severely underestimated using standard DFT functionals which plays a key role in calculating the optical properties (Dielectric function, refraction, reflectivity, and loss function etc.) Therefore, in the present study, we attempt to study the electronic transitions, refractive index and reflectivity through the calculated optical spectra for the five solid oxidizers using recently developed TB-mBJ potential. In general optical properties can be described by means of complex dielectric function $\epsilon(\omega)$ = $\epsilon_1(\omega)$ + $i\epsilon_2(\omega)$, where $\epsilon_1(\omega)$, and $\epsilon_2(\omega)$ represent the dispersive and absorptive part of the complex dielectric function $\epsilon(\omega)$. The imaginary part of dielectric function $\epsilon_2(\omega)$ is obtained from the momentum matrix elements between the occupied and unoccupied wave functions within selection rules and the real part of $\epsilon(\omega)$ can be extracted from the Kramers-Kronig relationships using $\epsilon_2(\omega)$.

\begin{figure}
\centering
\includegraphics[height = 4.5in,width=5.5in]{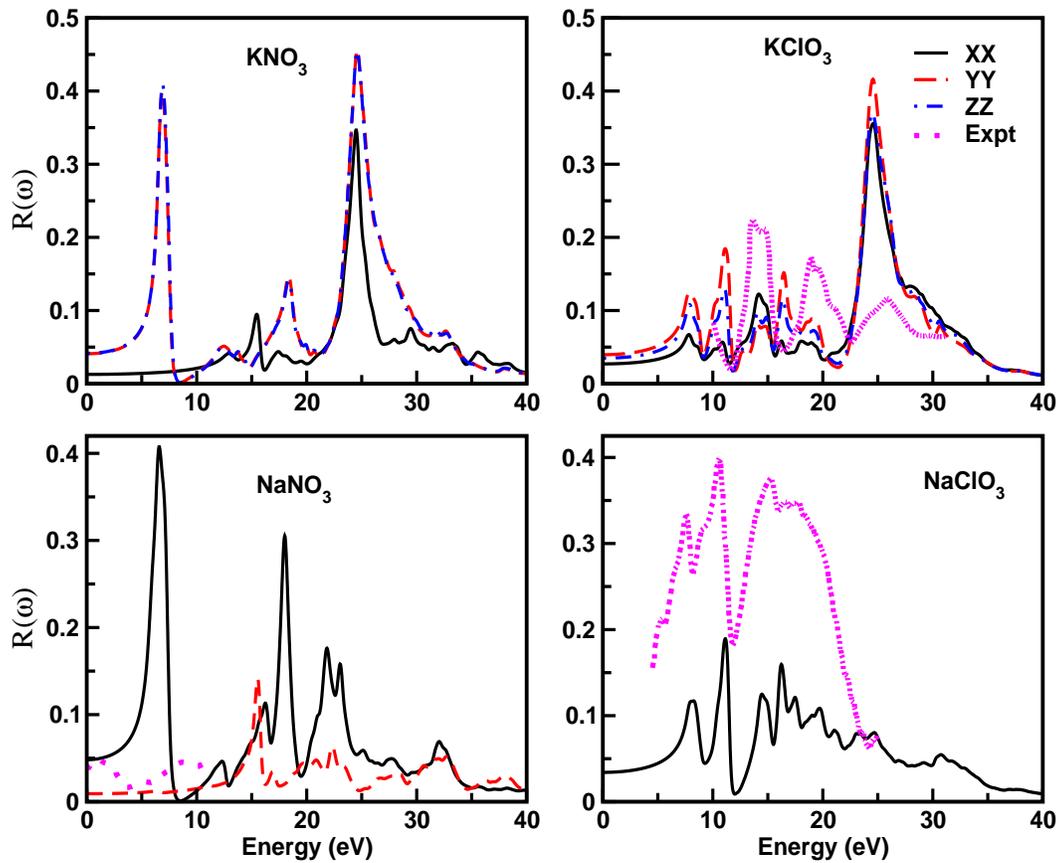}
\caption{\label{reflection} Calculated reflection spectra for NaNO$_3$, KNO$_3$, NaClO$_3$, and KClO$_3$ using TB-mBJ potential at the experimental crystal structures. The experimental data is taken from Refs.\cite{Vorobeva,Burkov} }
\label{reflection}
\end{figure}

As shown in figure \ref{Epsilon} the peaks (A, B, C, D, E, and F) in the $\epsilon_2(\omega)$ spectra along distinct crystallographic directions based on the crystal symmetry show that inter band transitions between occupied valence band and unoccupied conduction band. The optical transitions corresponding to each peak are given in Table 4 of the supplementary material for all the studied compounds. It is found that the optical transitions below 20 eV are due to anionic states or N/O/Cl $\rightarrow$ metal (Li, Na and K) atom whereas in the high (above 20 eV) energy region the transitions are due to cationic (K:p$\rightarrow$d) states, this can be clearly seen from figure \ref{Epsilon} for KNO$_3$ and KClO$_3$, and this is in very good agreement with quantum yield photo-emission measurements.\cite{Vorobeva} As illustrated in figure \ref{reflection}, the calculated reflectivity spectra reveal that the maximum reflection occurs at about 10 eV for (Li/Na)NO$_3$ and NaClO$_3$ while it is around 25 eV for potassium based oxidizers, and also the calculated spectra is consistent with the available experimental spectra at 9 K and 298 K.\cite{Vorobeva,Burkov} The calculated static refractive index are displayed in Table 5 of the supplementary material. A high value of refractive index represents the covalent bonding in a material, among the investigated compounds LiNO$_3$ possesses the highest refractive index value implying that it is more covalent than other studied compounds and refractive index values decrease from Li $\rightarrow$ Na $\rightarrow$ K (Na $\rightarrow$ K) in nitrates (chlorates) which shows that covalent (ionic) nature decreases (increases) in the same order for these materials. The calculated distinct static refractive index as a function of photon energy as shown in figure \ref{refraction} for the investigated compounds indicates that these materials are optically anisotropic crystals except NaClO$_3$ which is an isotropic crystal due to its cubic crystal symmetry and it is also found that above $\sim$ 20 eV these materials becomes optically isotropic up to maximum studied photon energy range.

\begin{figure}
\centering
\includegraphics[height = 4.5in,width=5.5in]{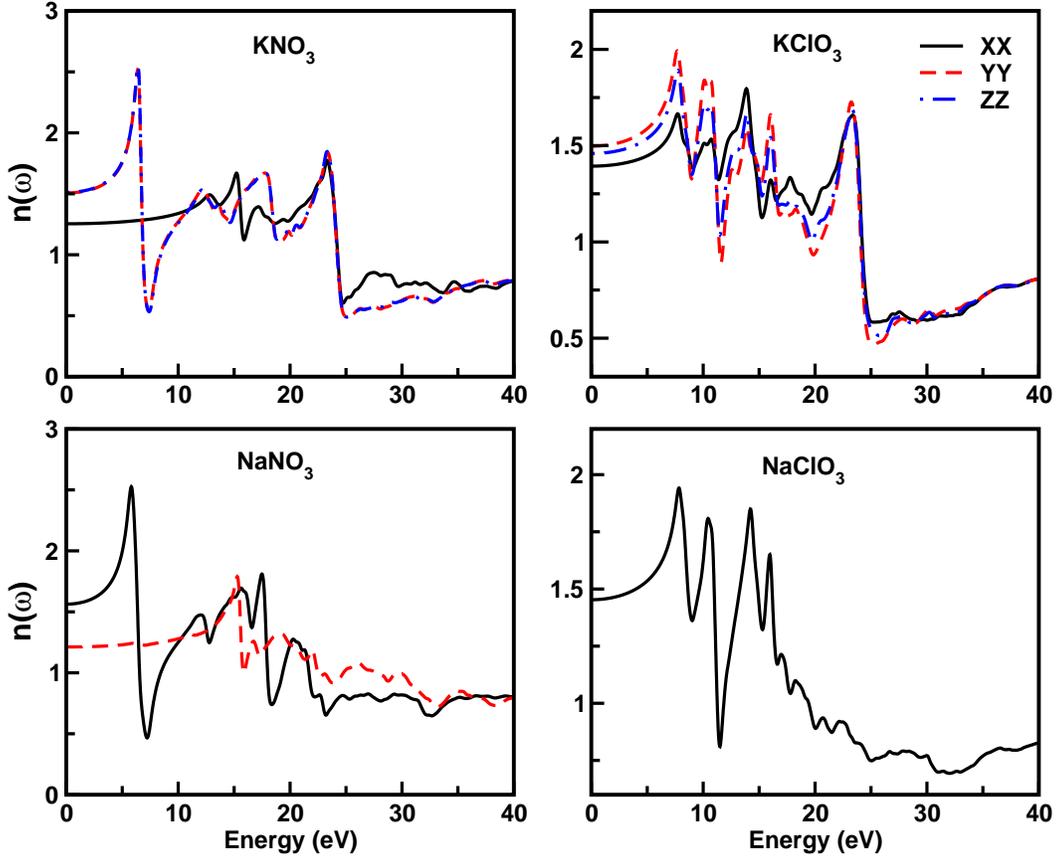}
\caption{\label{refraction} Calculated refractive index for NaNO$_3$, KNO$_3$, NaClO$_3$, and KClO$_3$ using TB-mBJ potential at the experimental crystal structures. }
\label{refraction}
\end{figure}

\section{CONCLUSIONS}
In summary, ab-initio calculations have been performed to investigate the structural, elastic, vibrational, electronic, and optical properties of solid oxidizers MNO$_3$ (M = Li, Na, K) and MClO$_3$ (M = Na, K) based on DFT. The calculated structural parameters are in good agreement with the experimental data in contrast to LDA and/or GGA results which show a large deviation from the experiments due to their inability in treating the weak dispersive interactions in ionic-molecular solids. Moreover, this method is successful in predicting phase stability of KNO$_3$ polymorphs. The calculated elastic constants at DFT-D2 equilibrium volume show that these materials are mechanically stable at ambient pressure. The obtained elastic and bulk moduli for the investigated compounds decrease in the following order: Li $\textgreater$ Na $\textgreater$ K (Na $\textgreater$ K) for nitrates (chlorates) due to increase in ionic size of the metal atom resulting low values of the elastic moduli indicating soft nature of the potassium-based compounds. We have calculated zone centre phonon frequencies and made a detailed vibrational assignment for the internal and lattice modes. We also noticed that the presence of metal (Li, Na , K) cation with nitrate/chlorate anion leads to an increase in the internal degrees of freedom in a crystal. From the calculated IR and Raman spectra, it is found that the vibrational frequencies show red shift from Li $\rightarrow$ Na $\rightarrow$ K (Na $\rightarrow$ K) and N $\rightarrow$ Cl for nitrates (chlorates) due to increase in mass of the metal and non-metal atoms, respectively. We have also calculated the electronic structure and optical properties of these materials using TB-mBJ potential and the calculated band gaps are found to improve over LDA/GGA functionals. The calculated electronic structure shows that these compounds are wide band gap insulators. Predominant ionic bonding between metal cation and nitrate or chlorate anion and covalent bonding (N-O/Cl-O) is present within nitrate/chlorate anion from the calculated PDOS which is consistent with the previous reports on these materials. We have also analysed electric-dipole transitions between valence band and conduction band, which are due to nitrate/chlorate group below 20 eV and cationic transitions occur above 20 eV from the calculated optical spectra and the calculated reflectivity spectra is consistent with the experimental data. We have also observed that these are optically anisotropic materials (except NaClO$_3$) and their anisotropic nature decreases in the high energy region.  \\

\subsection{ACKNOWLEDGMENTS}
NYK would like to thank DRDO through ACRHEM for financial support, and the CMSD, University of Hyderabad, for providing computational facilities. NYK acknowledges Prof. C. S. Sunandana, School of Physics, University of Hyderabad for critical reading of the manuscript.  \\


\clearpage
\begin{thebibliography}{00}

\bibitem{Hofmann}
Hofmann, A., et al, US Pat./ent, US20050072502 {\bf2005} A1.

\bibitem{Koch}
Koch, E. C., Propell. Explos. Pyrotech., {\bf2004}, 29, 67-80.

\bibitem{Venkatachalam}
Venkatachalam, S., Santhosh, G., Ninan, K. N., Propell. Explos. Pyrotech. {\bf2004}, 29, 178-187.

\bibitem{Meyer}
Meyer, R., Kohler, J., Homburg, A., Explosives, Wiley-VCH, Weinheim, {\bf2007}.

\bibitem{Vandel}
Vandel, A. P., Lobanova, A. A., Loginova, V. S., Russ. J. Appl. Chem., {\bf2009}, 82, 1763-1768.

\bibitem{Steinhauser}
Steinhauser, G., Klapotke, T. M., Angew. Chem. Int. Ed. {\bf2008}, 47, 3330-3347.

\bibitem{Agrawal}
Agrawal, J. P., High Energy Materials: Propellants, Explosives and Pyrotechnics, Wiley-VCH, Weinheim, {\bf2010}.

\bibitem{Sutton}
Sutton, G. P., Biblarz, O., Rocket Propulsion Elements, John Wiley Sons Inc., New Jersey, {\bf2010}.

\bibitem{Kubota}
Kubota, N., Propellants and Explosives: Thermo-chemical Aspects of Combustion, Wiley-VCH, Weinheim, {\bf2007}.

%

\bibitem{Witko}
Witko, E. M., Buchanan, W. D., Korter, T. M., J. Phys. Chem. A, {\bf2011}, 115, 12410-12418.

\bibitem{Fu}
Fu, X. , Song, Y., Sun, Ch., Zhou, J., Phys. Chem. Chem. Phys., {\bf2014}, 16, 6963-6967.

\bibitem{Johnson}
Johnson, M. C., Walker, D. , Clark, S. M., Jones, R. L., Am. Mineral., {\bf2001}, 86, 1367-1379.

\bibitem{Babar}
Babar, Z.-U.-D., Malik, A. Q., Caspian J. Appl. Sci. Res., {\bf2013}, 2, 63-69.

\bibitem{Liao}
Liao, L.-Q., Yan, Q.-L., Zheng, Y., Song, Z.-W., Li, J.-Q., Liu, P., Ind. J. Eng. Mater. Sci., {\bf2011}, 18, 393-398.

\bibitem{Dong}
Dong, X. F., Yan, Q. L., Zhang, X. H., Cao, D. L., Xuan, C. L., J. Anal. Appl. Pyrolysis, {\bf2012}, 93, 160-164.

\bibitem{Park}
Park, C. D., Choi, W. Y., Kang, H., Radiation Effects and Deffects in Solids, {\bf1990}, 115, 65-72.

\bibitem{Wu}
Wu, X., Fronczek, F. R., Butler, L. G., Inorg. Chem., {\bf1994}, 33, 1363-1365.

\bibitem{Gonschorek}
Gonschorek, W., Schmahl, W. W., Weitzel, H. , Miehe, G., Fuess, H., Z. Kristallogr., {\bf1995}, 210, 843-849.

\bibitem{Nimmo}
Nimmo, J. K., Lucas, B. W., J. Phys. C, {\bf1973}, 6, 201-211.

\bibitem{Adiwidjaja}
Adiwidjaja, G., Pohl, D., Acta Cryst. C, {\bf2003}, 59, i139-i140.

\bibitem{Abrahams}
Abrahams, S. C., Bernstein, J. L., Acta Cryst. B, {\bf1977}, 33, 3601-3604.

\bibitem{Danielsen}
Danielsen, J., Hazel, A., Larsen, F. K., Acta cryst. B, {\bf1981}, 37, 913-915.

\bibitem{Haussuhl}
Hauss\"uhl, S., Z. Kirstllogr, {\bf1990}, 190, 111-126.

\bibitem{Michard}
Michard, F., Plicque, F., Comput. Rend. B, {\bf1971}, 272, 848 and 1159.

\bibitem{Viswanathan}
Viswanathan, R., J. Appl. Phys., {\bf1966}, 37, 884-886.

\bibitem{RamaRao}
Sundara Rama Rao, B., Proc. Ind. Acad. Sci. A, {\bf1944}, 19, 93-99.

\bibitem{Ramdas1}
Ramdas, A. K., Proc. Ind. Acad. Sci. A, {\bf1953}, 37, 441-450.

\bibitem{Khanna}
Khanna, R. K., Proc. Ind. Acad. Sci. A, {\bf1961}, 27, 489-495.

\bibitem{Nedungadi}
Nedungadi, T. M., Proc. Ind. Acad. Sci. A, {\bf1941}, 14, 242-258.

\bibitem{Ismail}
Ismail, Md. A., Jayasooriya, U. A., Kettle, S. F. A., J. Chem. Phys., {\bf1983}, 79, 4459-4462.

\bibitem{French}
French, R., Devlin, J. P., Chem. Phys. Lett., {\bf1976}, 38, 79-82.

\bibitem{Ferraro}
Ferraro, J. R., Walker, A., J. Chem. Phys., {\bf1965}, 42, 1273-1277.

\bibitem{Brooker}
Brooker, M. H., J. Phys. Chem. Solids, {\bf1977}, 39, 657-667.

\bibitem{kumari}
Kumari, C. S., Proc. Indian Acad. Sci. A, {\bf1950}, 32, 177-183.

\bibitem{Ramdas2}
a) Ramdas, A. K., Proc. Ind. Acad. Sci. A, {\bf1952}, 35, 249-254. b) Ramdas, A. K., Proc. Ind. Acad. Sci. A, {\bf1952}, 36, 55-60. c) Ramdas, A. K., Proc. Ind. Acad. Sci. A, {\bf1953}, 37, 451-457.

\bibitem{Nakagawa}
Nakagawa, I., C.S.C Walter, J. L.,  J. Chem. Phys., {\bf1969}, 51, 1389-1397.

\bibitem{James}
James, D. W., Leong, W. H., J. Chem. Phys., {\bf1968}, 49, 5089-5096.

\bibitem{Rousseau}
Rousseau, D. L., Miller, R. E., Leroi, G. E., J. Chem. Phys., {\bf1968}, 48, 3409-3413.

\bibitem{Liu}
Liu, D., Ullman, F. G., Hardy, J. R., Phys. Rev. B, {\bf1992}, 45, 2142-2147.

\bibitem{Brooker1}
Brooker, M. H., Shapter, J. H., Drover, K., J. Phys.: Condens. Matter, {\bf1990}, 2, 2259-2272.

\bibitem{Hartwig}
Hartwig, C. M., Rousseau, D. L., Porto, S. P., Phys. Rev., {\bf1969}, 188, 1328-1335.

\bibitem{Bates}
Bates, J. B., J. Chem. Phys., {\bf1971}, 55, 494-503.

\bibitem{Brooker2}
Brooker, M. H., Shapter, J. G., J. Phys. Chem. Solids, {\bf1989}, 50, 1087-1094.

\bibitem{Heyns}
Heyns, A. M., Clark, J. B., J. Raman Spectrosc., {\bf1983}, 14, 342-346.

\bibitem{Neufeld}
Neufeld, J. D., Andermann, G., J. Phys. Chem. Solids, {\bf1973}, 34, 1993-2202.

\bibitem{Vorobeva}
Vorob\'eva, E. A., Kozhevnikov, A. V., Timchenko, N. A., Sheevtsov, A. A., Nucl. Instr. Meth. Phys. Res. A, {\bf1989}, 282, 615-618.

\bibitem{Zhuravlvev}
Zhuravlev, Y. N., Kravchenko, N. G., Poplavnoi, A. S., Dzyubenko, F. A., Spectroscopy, {\bf2002}, 92, 185-189.

\bibitem{Burkov}
Burkov, V. I., Makhov, V. N., Crystallography reports, {\bf2010}, 55, 272-275.

\bibitem{Aydinol}
Aydinol, M. K., Mantese, J. V., Alpay, S. P., J. Phys.: Condens. Matter, {\bf2007}, 19, 496210-1-496210-23.

\bibitem{Belomestnykh}
Belomestnykh, V. N., Sukhushin, Yu. N., Soviet Phys. J., {\bf1974}, 17, 56-58.

\bibitem{Zhuravlev1}
Zhuravlev, Y. N., Poplavnoi, A. S., Russ. Phys. J., {\bf2001}, 44, 391-397.

\bibitem{Zhuravlev2}
Zhuravlev, Y. N., Poplavnoi, A. S., Russ. Phys. J., {\bf2002}, 45, 1130-1131.

\bibitem{Zhuravlev3}
Zhuravlev, Y. N., Poplavnoi, A. S., Russ. Phys. J., {\bf2009}, 52, 881-883.

\bibitem{Zhuravlev4}
Zhuravlev, Y. N., Korabel\'nikov, D. V., Russ. Phys. J., {\bf2009}, 52, 965-970.

\bibitem{Zhuravlev5}
Korabel\'nikov, D. V., Zhuravlev, Y. N., J. Surf. Investig-X-ray Synchro., {\bf2013}, 7, 1067-1071.

\bibitem{Lovvik}
Lovvik, O. M., Jensen, T. L., Moxnes, J. F., Swarg, O., Unneberg, E., Comput. Mat. Sci., {\bf2010}, 50, 356-362.

\bibitem{Pradeep}
Pradeep, J., J. Int. Acad. Phys. Sci., {\bf2011}, 15, 337-344.

\bibitem{Payne}
Payne, M. C., Teter, M. P., Allen, D. C., Arias, T. A., Joannopoulos, J. D., Rev. mod. Phys, {\bf1992}, 64, 1045-1097.

\bibitem{Vanderbilt}
Vanderbilt, D., Phys. Rev. B, {\bf1990}, 41, 7892-7895.

\bibitem{Troullier}
Troullier, N., Martins, J. L., Phys. Rev. B, {\bf1991}, 43, 1993-2006.

\bibitem{Ceperley}
Ceperley, D. M., Alder, B. J., Phys. Rev. Lett., {\bf1980}, 45, 566-569.

\bibitem{Perdew}
J. P. Perdew, and A. Zunger, \emph{Phys. Rev. B}, 1981, {\bf23}, 5048-5079.

\bibitem{Burke}
Perdew, J. P., Burke, S. , Ernzerhof, M., Phys. Rev. Lett., {\bf1996}, 77, 3865-3868.

\bibitem{Almlof}
Fischer, T. H., Almlof, J., J. Phys. Chem., {\bf1992}, 96, 9768-9774.

\bibitem{Monkhorst}
Monkhorst, H. J., Pack, J. D., Phys. Rev. B, {\bf1976}, 13, 5188-5192.

\bibitem{Grimme}
Grimme, S., J. Comput. Chem., {\bf2006}, 27, 1787-1799.

\bibitem{heyd}
Heyd, J. Peralta, J. E., Scuseria, G. E., Martin, R. L., J. Chem. Phys. {\bf2005}, 123, 174101-1-174101-8.

\bibitem{perdew2}
Perdew, J. P., Parr, R. G., Levy, M., Balduz, J. L., Phys. Rev. Lett., {\bf1982}, 49, 1691-1694.

\bibitem{sham}
Sham, L. J., Schluter, M., Phys. Rev. Lett., {\bf1983}, 51, 1888-1891.

\bibitem{engel}
Engel, E., Vosko, S. H., Phys. Rev. B, {\bf1993}, 47, 13164-13174.

\bibitem{peter2}
Tran, F., Blaha, P., Phys. Rev. Lett. {\bf2009}, 102, 226401-1-226401-4.

\bibitem{becke}
Becke, A. D., Johnson, E. R., J. Chem. Phys. {\bf2006}, 124, 221101-1-221101-4.

\bibitem{koller1}
Koller, D., Tran, F., Blaha, P., Phys. Rev. B {\bf 2012}, 85, 155109-1-155109-7.

\bibitem{koller2}
Koller, D., Tran, F., Blaha, P., Phys. Rev. B {\bf 2011}, 83, 195134-1-195134-9.

\bibitem{singh2}
Singh, D. J., Phys. Rev. B, {\bf2010}, 82, 205102-1-205102-9.

\bibitem{camargo}
Camargo-Mart\'nez, J. A., Baquero, R., Phys. Rev. B, {\bf2012}, 86, 195106-1-195106-8.

\bibitem{jiang}
Jiang, H., J. Chem. Phys., {\bf2013}, 138, 134115-1-134115-7.

\bibitem{dixit}
Dixit, H., Saniz, R., Cottenier, S., Lamoen, D., Partoens, B., J. Phys.: Condens. Matter, {\bf2012}, 24, 2055030-1-2055030-9.

\bibitem{blaha}
Blaha, P, Schwarz, K., Madsen, G. K. H., Kvasnicka, D., Luitz, J., WIEN2K, an Augmented Plane Wave + Local Orbitals Program for Calculating Crystal Properties, Techn. Universitat: Wien, Austria. ISBN: 3-9501031-1-1-2, {\bf2001}.

\bibitem{konda}
Vaitheeswaran, G., Ramesh Babu, K., Yedukondalu, N., Appalakondaiah, S., CURRENT SCIENCE, {\bf2014}, 106, 1219-1223.

\bibitem{ramesh1}
Ramesh Babu, K., Vaitheesawaran, G., Chem. Phys. Lett., {\bf2012}, 533, 35-39.

\bibitem{ramesh2}
Ramesh Babu, K., Vaitheesawaran, G., J. Solid State Sci., {\bf2013}, 23, 17-25.

\bibitem{ramesh3}
Ramesh Babu, K., Vaitheesawaran, G., Chem. Phys. Lett., {\bf2013}, 586, 44-50.

\bibitem{Zhang}
Zhang, F., Gale, J. D., Uberuaga, B. P., Stanek, C. R., Marks, N. A., Phys. Rev. B, {\bf2013}, 88, 054112-1-054112-7.

\bibitem{Nimmo2}
Nimmo, J. K., Lucas, B. W. Acta Cryst. B, {\bf1976}, 32, 1968-1972,

\bibitem{Nimmo3}
Nimmo, J. K., Lucas, B. W., Nature, {\bf1972}, 237, 61-63,

\bibitem{Lu}
Lu, H. M., Hardy, J. R., Phys. Rev. B, {\bf1991}, 44, 7215-7224.

\bibitem{Mehl}
Mehl, M. J., Osburn, J. E., Papaconstantopoulus, D. A., Klein, B. M., Phys. Rev. B, {\bf1990}, 41, 10311-10323.

\bibitem{Mort}
Mort, K. A., Wilde, P. J., Jackson, R. A., J. Phys.: Condens. Matter, {\bf1999}, 11, 3967-3972.

\bibitem{Srivastava}
Srivastava, S. K., Sharma, S. K., J. Phys. Chem. Solids, {\bf2007}, 68, 1648-1651.

\bibitem{Born}
Born, M.; Huang, K.; Dynamical Theory of Crystal Lattices, Oxford University Press, Oxford, {\bf1998}.

\bibitem{Haussuhl1}
Hauss\"uhl, S., Z. Phys., {\bf1960}, 159, 223-229.

\bibitem{kondal}
Yedukondalu, N., Vikas Ghule, D., Vaitheeswaran, G., J. Chem. Phys., {\bf2013}, 138, 174701-1-174701-8.

\bibitem{Guravlev}
Guravlev, Y., Poplavnoy, V., Izv.Vuzov. Phys., {\bf2001}, 4, 51-54.

\end{thebibliography}
\end{document}